%% file: neurips-whitepaper.tex
\documentclass{article}

\usepackage[nonatbib,final]{neurips_2020}

\usepackage[utf8]{inputenc} 
\usepackage[T1]{fontenc}    
\usepackage{hyperref}       
\usepackage{url}            
\usepackage{booktabs}       
\usepackage{amsfonts}       
\usepackage{nicefrac}       
\usepackage{microtype}      

\usepackage{soul}
\usepackage{url}
\usepackage[utf8]{inputenc}
\usepackage{graphicx}
\usepackage{amsmath}
\usepackage{booktabs}
\newcommand{\clai}{{\tt CLAI}\xspace}
\newcommand{\nlcmd}{{\tt NLC2CMD}\xspace}

\usepackage{fancyvrb}
\usepackage{xcolor}
\usepackage{xspace}
\usepackage{mathtools}
\usepackage{amssymb}
\usepackage{booktabs}

\newcommand{\numresponses}{$235$}

\usepackage{etoolbox}
\usepackage{wrapfig}

\usepackage{multirow}
\usepackage{subcaption}

\usepackage{sidecap}

\title{
Project CLAI: Instrumenting the Command Line 
as a New Environment for AI Agents
}

\author{
Mayank Agarwal$^\dag$ $\cdot$ Jorge J. Barroso$^\dag$ $\cdot$ Tathagata Chakraborti$^\dag$ $\cdot$ Eli M. Dow$^\dag$\\[1ex] {\bf Kshitij Fadnis$^\dag$ $\cdot$ Borja Godoy$^\dag$ $\cdot$ Madhavan Pallan$^\ddag$ $\cdot$ Kartik Talamadupula$^\dag$ }
\AND
\textnormal{$^\dag$IBM Research}\\[1ex]
\textnormal{$^\ddag$ Individual Contributor}
}

\begin{document}

\maketitle

\begin{abstract}
This paper reports on Project \clai\ ({\bf C}ommand {\bf L}ine {\bf AI}) 
which aims to bring the power of AI to the command line interface (CLI).
The \clai platform sets up the CLI as a new environment for AI researchers 
to conquer by surfacing the command line as a generic environment that
researchers can interface to using a simple sense-act API, much
like the traditional AI agent architecture.
In this paper, we discuss the design and implementation of the platform in detail, through illustrative use cases of new end user interaction patterns enabled by this design, and through quantitative evaluation of the system footprint of a \clai-enabled terminal. We also report on some early user feedback on \clai's features from an internal survey.
\end{abstract}

\vspace{-10pt}
\section{The command line is back!}

For decades, the AI community has pursued the vision of autonomous 
assistants that operate with end-users inside computing systems. 
A key factor behind the stagnation of progress on this vision has been 
that AI developers and researchers -- who would together be tasked with bringing 
cutting-edge AI technology to such bots -- do not want to engage with the 
deep intricacies of the typical computing (operating) system. However, with 
the arrival of cloud-based ecosystems and cloud-native applications, as well
as the scalable real-world deployment of AI techniques, we are at an 
inflection point akin to the initial emergence of large-scale networked 
terminals. This is an opportune moment to transform the typical user's 
experience of computing systems, and imbue it with the power of AI.

One of the most powerful tools in software development is the 
command line, due to its speed and expressiveness.
However, another reason for the popularity of CLIs is that oftentimes, 
users {\em have to} use them.
This is proved by recent trends in software
development: GUIs can rarely keep up with the rate of change of features
(e.g. consider the time it took to move from Docker to
Kubernetes to OpenShift on cloud platforms).
This means that CLIs become the default interfacing medium not just for new adopters of a software, but also for experts in one domain
(e.g. programming) who are no longer experts in others (e.g. devops). 
The merging of developer and devops roles is certainly an emerging 
trend with the proliferation of cloud applications.
This is highlighted by newly emergent CLIs with smart features,
such as devspace \cite{devspace}, odo \cite{odo},
and so on for cloud native applications; 
and the well documented second coming of CLIs in popular media:
\cite{evangelist,avc,zdnet,zsh}.

\vspace{-5pt}
\paragraph{Why command line AI?}

However, even with the CLI's re-emergence, the issue of support
on the command line remains a huge problem. 
\cite{growth} shows the increasing complexity of CLI commands, 
while \cite{zombie} shows how community-sourced support has failed
to keep up with the needs of users, thereby motivating 
the need for on-premise support (such as CLI plugins)
with easy accessibility (natural language).
{\em Indeed, a bit of introspection in the support community \cite{zombie}
reveals that the reasons that community-sourced support
is failing -- e.g. overwhelming numbers of duplicate questions (that 
can be used for training), or straightforward questions (that can be 
answered from documentation), and so on -- are the same reasons that 
the state-of-the-art in AI can succeed.}

It is this very lacuna in the space of intelligent assistance and automation that Project \clai\ seeks to fill. 
Given the CLI as the fixed interface that users must utilize; and current advances in AI, ML, and NLP technology; a toolkit that combines them both is the need of the hour.
The software toolkit provides the developer with a heavily
instrumented version of Bash, the most commonly used command line
interface -- this acts as a whole new environment for the AI researcher
to conquer.






\subsection{Project \clai: A Challenge to the AI community}
\label{sec:challenge}

The AI community has always had a soft spot for AI assistants on the 
command line.
In the early to mid 90s, researchers at the University of Washington conducted extensive academic work in this space under the umbrella of 
``Internet Softbots'' \cite{etzioni1993planning,etzioni1994softbot}. 
These were AI agents that used a UNIX shell and the web to ``interact with 
a wide range of internet sources''.  The softbots provided a number of
novel features, including one of the first alternative interfaces to the
UNIX command line: a deliberative agent that could reason about multi-step
sequences of commands; and the ability to gather information about an open 
world. In the late 90s, Microsoft introduced a slew of assistive agents 
along with their Office productivity software. Unfortunately, the
(in)famous Clippy and other commercial softbots fell short of user 
expectations. Notably, that generation of embodied assistants taught 
future designers valuable lessons \cite{cooper2004inmates} for the
deployment of similar agents going forward.

More recently, a number of rule-based command line assistants such as 
{\tt bash-it},
{\tt oh-my-zsh},
{\tt thefuck},
{\tt tldr},
etc. have emerged. These CLI assistants generally deal with correcting misspelled commands and other common errors on the command line, as well as automating commonly performed tasks using predefined scripts. While these assistants certainly make the job of working with the command line easier, they have a high maintenance burden due to the constant up-keep of the rules that form their back-end. 
In general, such domain specific solutions do not scale or generalize; 
and recent advances in machine learning
\cite{li2019smartshell,lin2017program,lin2018nl2bash}
can make big contributions in this area.
The Linux Plan Corpus \cite{linux-plan-corpus}
-- a collection of Linux sessions -- has become a great source of data for research in this direction \cite{blaylock2004statistical}.
Our own prior work on the UbuntuWorld system \cite{chakraborti2017ubuntuworld} used 
a combination of automated planning, reinforcement learning (RL), and 
information retrieval to drive data-driven exploration and decision making 
on the CLI, by bootstrapping data from the AskUbuntu forum. 
Similarly, researchers have also attempted to use RL for interpreting actionable commands from natural language instructions in Windows \cite{branavan2009reinforcement}.
With recent advances in AI, especially in machine learning and natural 
language processing, learning agents on the command line are poised for 
a comeback
\cite{lin2018nl2bash,betty,gros2019ainix,lin2017program,li2019smartshell}. 

\subsubsection{Cross-disciplinary challenges}

Interestingly, the command line environment provides a suite of challenges 
to the broader AI community as well: this includes multi-agent systems (in 
orchestration of multiple plugins on the platform); and automated planning and 
reinforcement learning (in being able to observe user behavior and 
learn over continuous interactions). 
Most importantly, it calls for cross-collaboration with the broader human-computer interaction community so that the mistakes of the past
\cite{cooper2004inmates} are not repeated.

\vspace{-2mm}
\subsubsection{Challenge to the State of the Art}

\paragraph{Energy Footprint.}

As we highlight later in our user study: 
{\em .. the responses from bots residing on the CLI
must be instantaneous, and not allow for any noticeable lag.
More than 75\% indicated that they expected a
comeback within 3 seconds. 
Furthermore, these skills cannot consume too much power 
(computational or otherwise), since 
they must ultimately run on (end) terminals.} 
As has been documented frequently, the state-of-the-art 
in machine learning largely ignores the real-world cost
of running AI models~\cite{meena,strubell2019energy}.
The \clai platform surfaces this as a core challenge to the community.

\vspace{-5pt}
\paragraph{Neurosymbolic AI.}

Another emerging theme in AI is the unification of end-to-end data-only 
approaches with those that operate on knowledge. 
For command line support, this is going to be {\em necessary} since it is 
impossible to generate unseen commands from the data in (for example) 
Stack Overflow alone.
However, the domain does come with a lot of structured knowledge, such as man (manual) pages; and forces the skill developer to 
figure out the best synergy of both worlds.
\clai thus stands to give a big boost to
this emerging research theme.

\clearpage

\begin{figure}
\centering
\includegraphics[width=0.75\columnwidth]{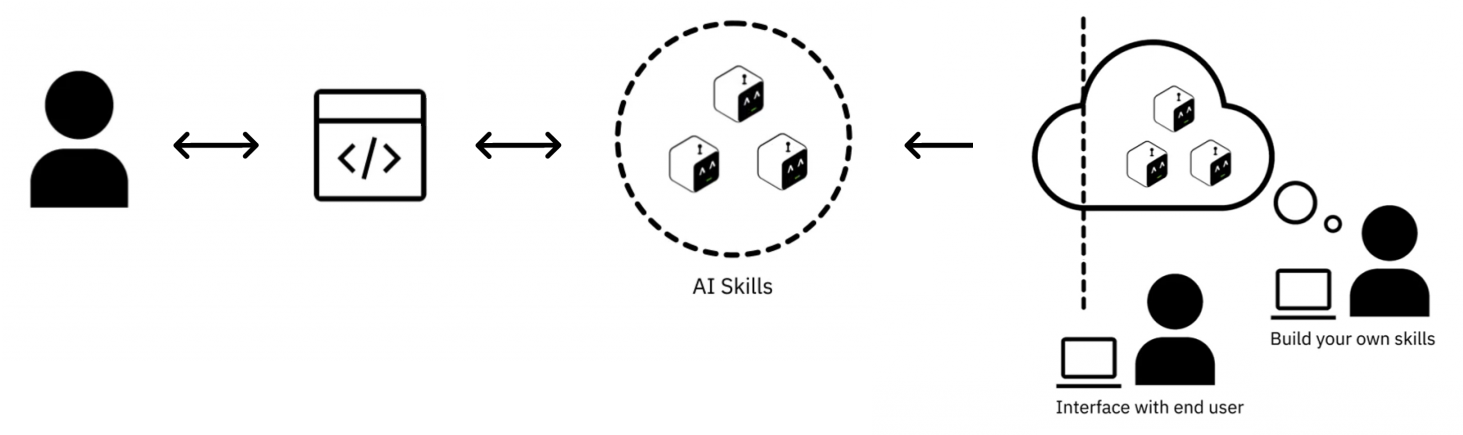}
\caption{
Overview of \clai: 
the developer, while working on the command line interface, has
access to an assistant in the form of a orchestrated set of skills.
The AI researcher has access to an API that gives them access to
an instrumented command line for building AI capabilities. 
}
\label{clai-big-idea}
\end{figure}



\section{\clai: System Overview}
\label{sec:clai_system_overview}

At the core of \clai are AI plugins or ``skills'' 
that monitor every user action on the terminal.\footnote{Although 
the architecture is generalizable to a large extent, we currently only support Bash since it is the most commonly used CLI. As a proof of 
this concept, we have recently ported all the discussed functionalities into the 
USS terminal in the z/OS operating system on mainframes.
Those results are not public yet.
In the rest of this paper, unless otherwise mentioned, we will thus be referring 
to Bash whenever we mention command line, terminal, or shell in the context of 
the \clai platform or infrastructure.} 
This is equivalent to the notion of skills in IBM's Watson Assistant
(\textcolor{blue}{\url{ibm.co/2LblJ70}}) 
or Amazon's Alexa 
(\textcolor{blue}{\url{amzn.to/2ZH9Olp}}) 
-- a skill is a function that performs microtasks. 
Every command that the user types, or any execution process on the terminal, 
is piped through the skills active in \clai for that session. 
A skill can thus autonomously respond to any event on the terminal. 

\subsection{\clai for the researcher: The \clai API}

An important user-persona of \clai is the
developer/researcher who creates the skills. 
\clai makes Bash available to a skill developer via a generic
``environment'' API, so that the developer does not have to deal with interfacing
issues and can instead focus on building their AI plugins. 
In order to make this very familiar to the AI community, 
this interface allows execution of actions and sensing 
of the result of those actions in a manner very similar to the classic AI agent
architecture \cite{russell2016artificial,sutton1998introduction}.
-- imagine that same classic AI agent architecture, but replace the 
environment with Bash. 
This API thus makes Bash available as another new, exciting playground 
for AI agents, much like OpenAI {\tt Gym}: \textcolor{blue}{\url{gym.openai.com/}}.
The \clai API -- built in {\tt Python3} -- has two major components.

\begin{wrapfigure}{R}{0.45\textwidth}
\vspace{-20pt}
\begin{center}
\includegraphics[width=0.44\columnwidth]{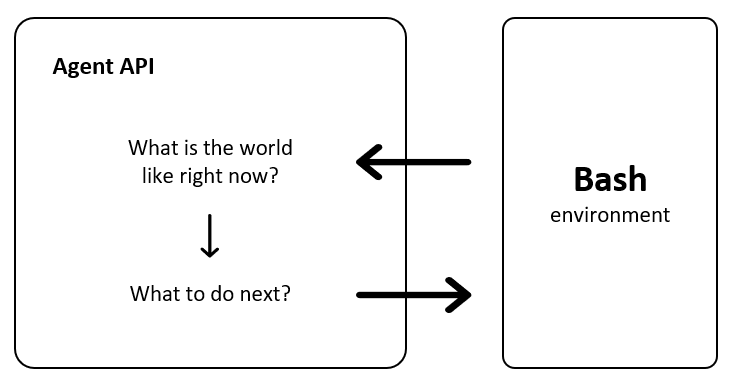}
\end{center}
\vspace{-10pt}
\end{wrapfigure}

\paragraph{The \clai Skill API}
This lets a developer intercept as well as execute a callback on every user input 
on the command line {\em after the user hits return}, and lets them respond
appropriately. Developers can:
(a) Do nothing and let normal life on the command line follow.
This includes doing nothing but registering an event to
learn from that event and/or track user state;
(b) Add something to the user input -- e.g. a flag that would make a command work;
(c) Replace the user input by something else -- e.g. respond to a natural 
language request for automation;
(d) Respond to the outcome (e.g. error) of a user command for in-situ support 
and troubleshooting;
(e) Add something to the outcome -- e.g. for proactive support or pedagogical
purposes.


\begin{itemize}

\item[{\bf State}] 
The State object contains information about 
the system -- including the state of the shell, system memory, connectivity, file system,
etc. -- as the state information or percept received from the terminal session that a
skill is plugged into. 

\item[{\bf Action}] 
The Action object is the directive from the skill to the terminal. 
This includes the suggested command, a description and an explanation
for the suggestion, and other control parameters that 
control the user action (such as permissions to execute).
The skill can also return a sequence of Action objects in response 
to a user command or terminal event to complete a process based on
the current user intent.

\end{itemize}

\paragraph{The \clai Orchestration API}

The \clai assistant is realized in the form of an orchestrated set
of skills -- all active skills communicate with the ``orchestration layer''
that decides whether to pass on their individual responses to the terminal.
The orchestrator enforces the following contract between the terminal
and the skills: every event is reported to all active skills 
in the form of the State Object, and every skill responds with 
an Action object (or a list of Action objects) with an associated
confidence (self-determined by the skill) of its relevance and/or accuracy.
This {\tt act-sense} cycle allows an AI agent plugged into Bash to act and 
learn over continued periods, either by itself \cite{chakraborti2017ubuntuworld} 
or in the course of prolonged interaction with the user. 

\subsection{\clai for the end user: Interaction Patterns}
\label{interaction-patterns}

\paragraph{\clai in the background.} 

\begin{wrapfigure}{r}{0.42\textwidth}
\vspace{-15pt}
\begin{Verbatim}[commandchars=\\\{\},fontsize=\small]
>> <command>
\textcolor{teal}{\clai: \textnormal{augment and / or replace} <command>}
\textcolor{blue}{user: y/n/e}
<stdout>
\textcolor{teal}{\clai: \textnormal{augment to} stdout}
<stderr>
\textcolor{teal}{\clai: \textnormal{respond to} stderr}
\end{Verbatim}
\vspace{-15pt}
\end{wrapfigure}

The other user-persona of \clai -- the actual user of the command line 
-- has three ways of using \clai skills.
In the first mode, the user's input most closely resembles 
normal life on Bash. For most commands, the user experience is entirely
unchanged.\footnote{This is a conscious design decision in light of lessons 
learned from historical deployments of assistants in operating systems,
often deemed to be unnecessarily obtrusive \cite{cooper2004inmates}.}
When a skill does get invoked, the user will experience the following 
interfacing pattern:


\vspace{-5pt}
\begin{itemize}

\item[-] \clai may replace the user input command (or augment it) in order to make
execution work as the user likely intended. Users see the augmented or altered input
command and may approve the input variant for execution, or ask for an explanation on
the command substitution rationale. This may be used to for example fix a mistake in
the command on the fly or translate a natural language input on the command 
line to their Bash syntax.

\item[-] \clai may add additional information to the {\tt stdout}.
This may be used, for example, for pedagogical purposes (e.g. a better
way to perform that task) or for alerting the user to certain system information,
or just enhancing functionality of existing Bash utilities.

\item[-] \clai may respond to the {\tt stderr} by
providing additional information for troubleshooting, or 
by suggesting a fix that the user can follow through on
in their next command. 

\end{itemize}

\vspace{-5pt}
\paragraph{\clai explicitly invoked.}

In order to force assistance from \clai, a user may opt to demand a response from
\clai using the syntax below. Doing so will cause \clai to respond with 
the skill that it believes to be most relevant to the context
{\em bypassing the determination of relevance from the orchestration layer}
(e.g. ignore low confidence of a response).

\vspace{-5pt}
\begin{Verbatim}[commandchars=\\\{\},fontsize=\small]
\textcolor{teal}{>> \clai <command>}
\end{Verbatim}

\vspace{-5pt}
\paragraph{\clai forced skill invocation.}

Finally, the user can force a particular skill to respond, 
{\em regardless of the orchestration pattern} (e.g. ignoring its 
confidence and those of other skills currently active):

\vspace{-5pt}
\begin{Verbatim}[commandchars=\\\{\},fontsize=\small]
\textcolor{teal}{>> \clai <skill_name> <command>}
\end{Verbatim}

\section{\clai in Action}
\label{action}

In this section, we provide details of skills
and orchestration patterns illustrating the various 
capabilities of \clai\ and typical user
interactions with an assistant on the command line. 
Please refer to the appendix
for screenshots of all the examples in the paper.

\subsection{\clai Skills}

\clai comes with a few default skills aimed to demonstrate 
a large set of features both to the end users as well
as potential skill developers. 
They fall into one or more of the following categories.

\subsubsection{Skill Categories}

Based on the interfacing options described in Section \ref{interaction-patterns}, 
a few key interaction patterns emerge. While these are not intended 
to be exhaustive, they do capture some of the most interesting 
interaction types that we have explored so far with \clai. 



\begin{wraptable}{r}{66mm}
{\small
\begin{tabular}{l|l}
\textcolor{blue}{Bash} & \textcolor{blue}{\clai-enabled Bash}\\
\textcolor{teal}{user leaves to figure out stuff} & >> do xyz\\
>> command p & \clai: command p? y/n/e \\
& user: y
\end{tabular}
}\vspace{-10pt}
\end{wraptable}

\paragraph{Natural language support.}

This pattern allows the user to interact with the command line in natural language:
e.g. the user can ask {\tt >> how do I extract file.bz2}, or tell the terminal to {\tt >> extract all images from this archive}.

\begin{wraptable}{r}{50mm}
\vspace{-15pt}
{\small
\begin{tabular}{l|l}
\textcolor{blue}{Bash} & \textcolor{blue}{\clai-enabled Bash}\\
>> command p & >> do task xyz\\
>> command q & \clai: command p\\
>> command r & \clai: command q\\
& \clai: command r
\end{tabular}
}
\vspace{-15pt}
\end{wraptable}

\vspace{-5pt}
\paragraph{Automation.}

This pattern allows the user to hand off complex tasks (achieved by one or more commands) to \clai -- e.g. deployment of an application to the cloud. 
Developers already write scripts to achieve some of this functionality. This feature rescues them from writing tedious automation scripts and instead provide task-dependent automation. The most obvious technology match is to established AI techniques such as automated planning and reinforcement learning. 


\begin{wraptable}{r}{66mm}
\vspace{-5pt}
{\small
\begin{tabular}{l|l}
\textcolor{blue}{Bash} & \textcolor{blue}{\clai-enabled Bash}\\
>> command q & >> command q\\
Error: xyz & Error: xyz\\
\textcolor{teal}{user leaves to figure out stuff}  & \clai: command p \\
>> command p & >> command p\\
>> command q & >> command q
\end{tabular}
}
\vspace{-5pt}
\end{wraptable}

\vspace{-5pt}
\paragraph{On-premise support.}

Currently, when the command line user encounters an error, the usual response is to
indulge in the following loop: copy the error from the terminal, go over to a web
browser, search on the internet, copy the top answer, and come back to the terminal 
to try it out.
This is a frustrating and repetitive pattern of interaction on CLIs.
The in-situ support and troubleshooting pattern of \clai brings help from online
forums, support communities, and support documentation, 
directly to the terminal, so that users do not have 
to remove themselves from their immediate work context. 
This ensures that the support can be 
(1) {\em local or personalized} to the user's system; 
(2) {\em immediate}; and (3) {\em in-situ} without the user losing context. 


\begin{wraptable}{r}{66mm}
\vspace{-5pt}
{\small
\begin{tabular}{l|l}
\textcolor{blue}{Bash} & \textcolor{blue}{\clai-enabled Bash}\\
>> command q & >> command q\\
Error: xyz & \clai: command p\\
\textcolor{teal}{user leaves to figure out xyz}  & \clai: command q\\
>> command p &\\
>> command q &
\end{tabular}
}
\vspace{-5pt}
\end{wraptable}

\vspace{-5pt}
\paragraph{Proactive support.}

In certain situations, \clai can anticipate errors and let the user know about those
errors (or even go ahead and fix them in the background) in advance. For example, it
could be the case that a user might need to free up space on a cloud instance before 
proceeding to deploy an application. In such cases, \clai skills can catch and
prevent future errors that the user would otherwise encounter on the standard 
command line.

\begin{wraptable}{r}{66mm}
\vspace{-5pt}
{\small
\begin{tabular}{l|l}
\textcolor{blue}{Bash} & \textcolor{blue}{\clai-enabled Bash}\\
>> command p & >> command p\\
stdout+stderr & stdout+stderr\\
& \clai: q may be a better option
\end{tabular}
}
\vspace{-5pt}
\end{wraptable}

\vspace{-5pt}
\paragraph{Pedagogy and Augmentation.}

The \clai system can also chime in from time to time and help the user with their proficiency on
the terminal. This could involve something as simple as letting them know about new
features (e.g. letting the user know that the new way of running Flask applications
is {\tt >> flask <file>} when they type in {\tt >> python <file>}); or in the long 
run even retaining and guiding (for example) a new adopter of cloud platforms into
becoming an expert on the cloud. 
This may also involve augmenting the standard functionalities of existing 
command line utilities with new capabilities by adding to the standard
output new information that may enhance the user experience.




\subsubsection{\clai Available Skill Catalog}

\paragraph{nlc2cmd}
This is the canonical example of a natural language
interface to the terminal.
It connects to a Watson Assistant (WA) instance in the backend
to interpret user intents, and translate those to 
popular bash commands like {\tt tar}, {\tt grep}, etc.
Each Bash utility here corresponds to a separate 
WA workspace, and each flag of that utility 
maps to a specific intent with the parameters being
the corresponding entities: thus each command is uniquely
mapped to a corresponding natural language understanding
layer given its man (manual) page. 

\vspace{-5pt}
\paragraph{tellina}
Manually transforming man pages into NLU workspaces does not scale.
This skill integrates the state of the art of the nlc2cmd use case: 
Tellina \cite{lin2018nl2bash}. 
Tellina builds on recent advances in the application of deep learning
techniques to natural language processing (NLP) by 
adopting Seq2Seq \cite{sutskever2014sequence} models
from translation from English to a set of command templates.
This model is augmented with an argument filling module that performs 
template generation followed by program slot filling in order to fill 
out the various slots in the command template correctly.

\vspace{-5pt}
\paragraph{\clai fixit}
This skill provides help in response to the last command executed, by 
echoing back the response from the massively popular {\tt thefuck} plugin
for the command line. This skill is meant to illustrate how to integrate 
existing Bash plugins into the \clai platform.

\vspace{-5pt}
\paragraph{man page explorer}
This skill interprets questions in natural language, 
and responds with the most relevant command it can find from
the man pages installed in the system.
It also augments its response with a concise 
description of the man page using the popular {\tt tldr} plugin. 
This is an illustration of {\em both natural language support
as well as plugin integration}.
This agent trains a {\tt scikit-learn} \cite{scikit-learn} 
TF-IDF vectorizer \cite{ramos2003using}
over the retrieved man pages, and uses the cosine similarity between the 
natural language question and the man page contents to suggest a command.

\vspace{-5pt}
\paragraph{\clai howdoi}
This is very similar to the man page explorer, but instead of using manuals, 
it responds with the most relevant
answer from Unix Stack Exchange \cite{stackexchange}.
The posts and their highest rated answers are indexed into 
Elasticsearch \cite{gormley2015elasticsearch}. 
The query is compared against this index and the most relevant post 
and its accepted answer is returned.

\vspace{-5pt}
\paragraph{\clai helpme}
This is identical to {\tt howdoi} except that it fires when there is 
a standard error.

All the above skills have the same purpose: to ensure that
users of the command line do not have to leave context 
every time they face an inane error or do not have the
syntax memorized, and go looking for answers on the internet.
The following skills, on the other hand, explore much more 
sophisticated use cases demonstrating how \clai expands
to way beyond episodic support and troubleshooting.

\vspace{-5pt}
\paragraph{gitbot}

This bot helps a developer navigate their
local git setup and their GitHub repository from the command 
line. It highlights two use cases: 1) the use of a local
natural language layer built on Rasa NLU \cite{rasa}
as opposed to calls to external servers in 
{\tt nlc2cmd} and {\tt tellina}\footnote{This 
design decision has serious consequences -- on the one 
hand local skills can adapt to individual users better and 
are preferred in terms of security and privacy and do not need
an internet connection to function on a machine. 
They are also faster.
However, it does come with much higher memory and compute footprint 
on the local system as well as significantly more installation overhead.}
-- this replicates the nl2cmd use case but 
on git commands; and 2) Illustration of the use of the GitHub
Actions API \cite{github} to control the online GitHub repository
(issues, pull requests, etc.) without leaving the command line. 


\vspace{-5pt}
\paragraph{\clai dataXplore}

Data science has become one of the most popular real-world 
applications of ML. 
The {\tt dataXplore} skill is targeted specifically toward making the CLI 
easier to adopt and navigate for such data scientists.
The current version of the skill provides two functionalities: {\tt summarize} and {\tt plot}. 
{\tt summarize} utilizes the {\em describe} function \cite{pandasDescribe} of
the popular Pandas library \cite{pandas} to generate a human-readable summary
of a specified {\tt CSV} file; this functionality is intended to allow data
scientists to quickly examine any data file right from the command line.
{\tt plot} builds on the {\em plot} function provided by MatPlotLib \cite{matplotlib}, and the {\em pillow} \cite{PillowPlot,PIL} library to generate a plot of a given {\tt CSV} file. 
Such functionalities illustrate how \clai\ can be used as a CLI assistant 
for data science. 



\vspace{-5pt}
\paragraph{cloudbot}
This is a stateful agent that can 
automate tasks involving Docker \cite{merkel2014docker} and Kubernetes \cite{brewer2015kubernetes}
requiring the execution of a sequence of 
actions by harnessing automated planning techniques
as an instance of the automation use case.
The role of the planner here is 
to generate scripts that would otherwise have 
to be specified manually 
-- for example, while deploying an application to the cloud. 
In addition to automating the lengthy deployment pipeline,
the {\tt YAML} file that currently needs to be written manually
is generated automatically by the skill by: 1) monitoring user activities
on the terminal; 2) pinging the cloud account for the types of services 
available; and 3) parsing the Dockerfile. 
The planner used is \cite{helmert2006fast}.
This skills also demonstrates integration of 
continuous monitoring of user
state and plan recognition to continuously predict a possible intent and 
re-plan as the user executes commands on their terminal. 
The plan recognizer used is \cite{ramirez2010probabilistic}.


\subsection{Orchestration Patterns}
\label{whale}

The orchestration layer comes with a unique set of challenges~\cite{Rizk2020AUnified}. 
There may be two approaches to orchestrate skills: 
1) {\bf apriori}, where the orchestrator acts as a filter and decides which plugin to invoke; and 2) {\bf posterior}, where all plugins listen and respond, and let the orchestrator pick the best response (this is the current setup).
The apriori option is likely to have a smaller system footprint, but
involves a single bottleneck based on the accuracy of the classifier which determines which plugin to invoke. 
Furthermore, this requires that the orchestrator be cognizant of the full list of plugins and their capabilities -- this is unrealistic.
The {\em posterior} option -- despite increased latency and computational load -- keeps the skill design process independent from the orchestration layer.
Skill confidences can be calibrated over time
by learning from user responses to \clai actions: either directly from their 
\texttt{y/n/e} responses; or indirectly by observing what command 
they executed after a suggestion, and matching that to how similar 
it is to the suggested course of action.

\vspace{-5pt}
\paragraph{Rule-based orchestration.}

Among the rule-based orchestration patterns packaged with the software are the following:
1) Max-orchestrator: this pools responses from the active skills 
and passes the response with the highest confidence, above a threshold
specified by either the user or the developer, on to the command line;
2) Threshold-orchestrator: This is the max-orchestrator where the threshold 
is bumped up or down based on user feedback; and
3) Preference-based orchestrator: The user can provide partially ordered
preferences above the thresholding-and-max mechanism. 

\vspace{-5pt}
\paragraph{Learning-based orchestration.}

With continued feedback from the end user, 
much more sophisticated orchestration patterns can be 
learned \cite{upadhyay2019bandit}.
This is especially useful in adapting the assistant 
to specific users and user types. \clai comes packaged with
a contextual bandit based orchestrator, which uses the confidences
returned by each skill as the context vector and decides
which skill should respond to the user command. The user 
feedback on the orchestration choices is used to
reward the contextual bandit model, which in turn helps it to 
adapt to the user's requirements. 
Additionally, to avoid the initial 
exploration phase which can adversely affect the user experience, 
the bandit model can be warm-started with a specific user profile.
We include four warm-start profiles: 
1) {\bf ignore-clai}: this warm-starts the orchestrator to
ignore all \clai responses; 
2) {\bf max-orchestrator}: this warm-starts the bandit 
to select the skill that responds with the maximum confidence; 
3) {\bf ignore-skill}: this warm-starts the bandit to ignore a 
particular skill and behave as a max-orchestrator otherwise; and
4) {\bf prefer-skill}: this warm-starts the bandit to prefer a particular 
skill over another and is useful in cases where the user has preferences over
skills with overlapping domains -- e.g. a user on a Mac terminal 
might not have use for a skill that retrieves data from the Unix Stack 
Exchange. 


\begin{figure}
\centering
\begin{minipage}{0.33\textwidth}
\centering
\includegraphics[width=\linewidth]{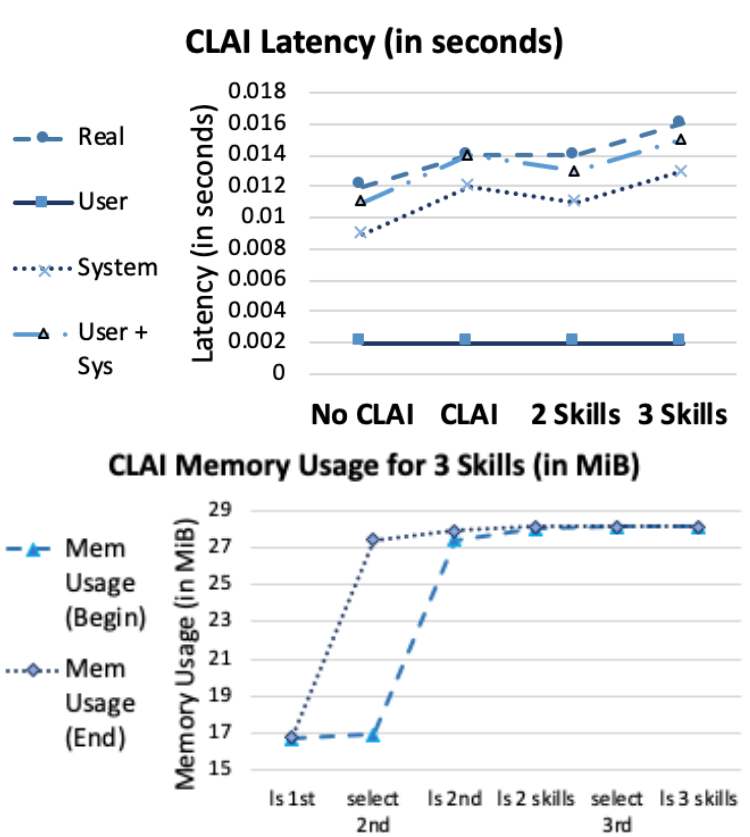}
\caption{\clai system footprint: (a) Latency; (b) Memory.}
\label{fig:bb-footprint-graph}
\end{minipage}%
\hfill
\begin{minipage}{0.6\textwidth}
\centering
\includegraphics[width=0.99\textwidth]{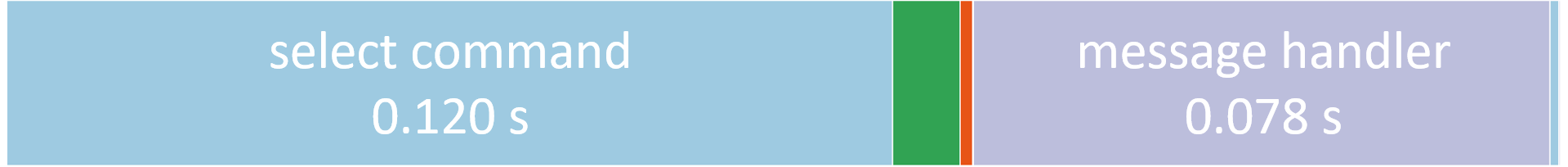}
\caption{
Latency profile of activating a \clai skill. This activates the \clai core along with the activation procedure of the specified skill, thus increasing the latency but still keeping it well below the preferred sub-second mark.
\vspace{15pt}
}
\label{fig:cla1-latency-tar-1}
\centering
\includegraphics[width=0.99\textwidth]{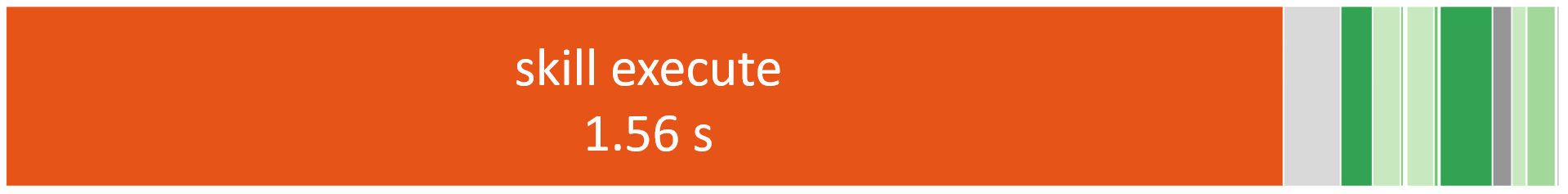}
\caption{
Latency profile of executing a bash command with installed \texttt{clai skills}. This activates the \clai core along with \clai skills. While the latency increases significantly to 2s, around 80\% of it is contributed by the skill itself.
}
\label{fig:cla1-latency-tar-2}
\end{minipage}
\end{figure}

\section{Internal Evaluations}
\label{yeti}

One of the primary challenges in deploying a framework like \clai
is ensuring that the resources consumed do not hinder the user experience 
on the command line. 
This is evident from our survey of end users -- a whopping $80\%$ of the 
respondents (and $93\%$ of developers/devops) require a latency of less 
than $3$ seconds, with more than half that number requiring an even more 
stringent sub-second latency.
This, in addition to the energy and compute footprint,
makes this domain especially challenging 
to the state of the art in AI. 
More details are provided in the appendix.

\vspace{-5pt}
\paragraph{System Footprint}

We profile \clai to understand the latency of the system  using \texttt{yappi} \cite{yappi}. Since \clai consists of the \clai core and pluggable
skills, we focus on two scenarios: 1) when only the \clai core is invoked; 
and 2) when the \clai core is invoked along with \clai skills. 
We perform our analysis on a Quad-Core Intel i7 processor with 
16 GB RAM, and observe that the core \clai functionality of listing \clai
skills takes only 0.05 seconds; 
while more computationally expensive core \clai 
functionality such as activating a skill takes about 0.2 seconds. 
On the other hand, using \clai with installed skills increases the 
latency to about $2$ seconds, with the skills contributing about 80\% to 
the latency. More details are presented in the appendix.

It is clear from the profiling results that the onus of making \clai 
more responsive rests on the skills. 
Figure \ref{fig:bb-footprint-graph} shows how the latency
varies with increasing numbers of skills.
Since each skill is executed in parallel threads,
there is little change in the user (client) time trend.
This is encouraging, since it suggests that the client side, with limited
ability to scale its compute,
shows near-constant latency. In most cases where the computation happens 
on the client side, skill developers
cannot rely on the availability of specialized hardware to accelerate their
skills, and instead would need to ensure faster skill inference procedures
on standard user machines.
Specifically for deep neural network 
based skills, faster inference methods on CPUs have been proposed
\cite{lane2016deepx, teerapittayanon2016branchynet, zen2016fast, zhang2018deepcpu}, and \clai provides a practical test bed for further 
avenues of research into these areas.


\vspace{-5pt}
\paragraph{User Evaluations}
\label{subsec:user_evaluations}


\begin{figure*}[tbp!]
\centering
\includegraphics[width=\textwidth]{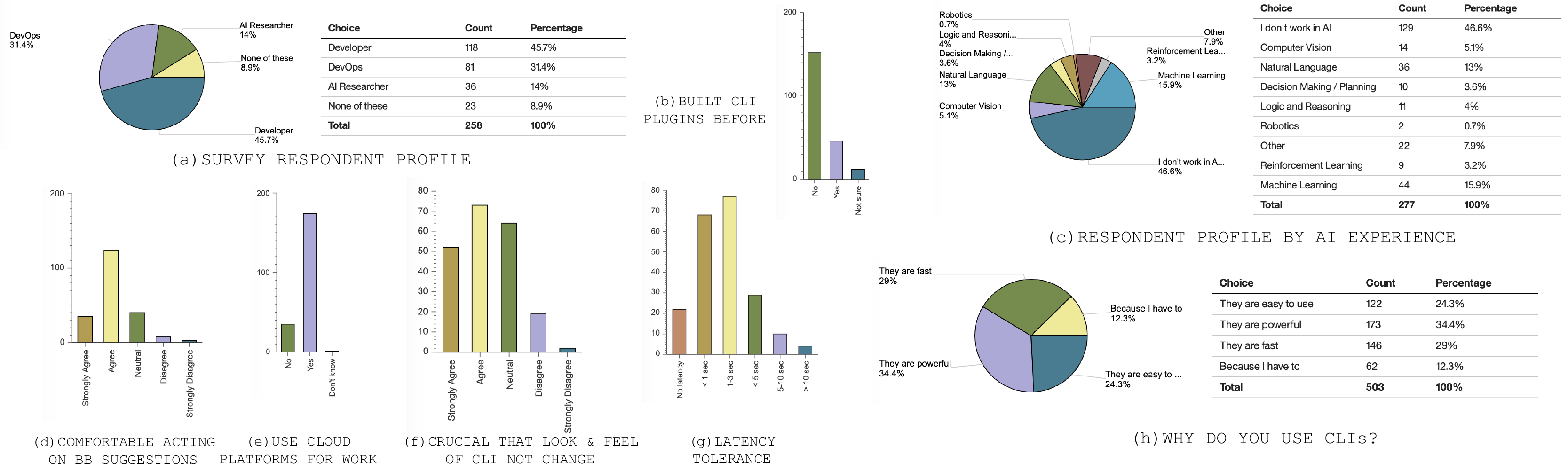}
\vspace{5pt}
\caption{Aggregate results from our user study on \clai. }
\label{fig:results-aggregate}
\end{figure*}

We  report on user feedback on \clai from an internal survey based on \numresponses\ responses.
Figure~\ref{fig:results-aggregate}(a) offers an {\bf aggregate profile} of the survey respondents.
More than three-quarters of the responses came from respondents who identified as either developers or devops, while only around $14\%$ of the respondents identified as AI practitioners. 
This 
indicates the potential for \clai to positively impact communities that have hitherto not had too much interaction with state-of-the-art AI techniques and technologies. 
Figure~\ref{fig:results-aggregate}(c) zooms in on the respondents' {\bf AI background}. Roughly over half of the respondents had some past AI experience. Of these, Machine Learning was the top area, with Natural Language (Processing) a close second. 
Figure~\ref{fig:results-aggregate}(e) shows that nearly $4$ out of $5$ respondents report {\bf usage of cloud platforms} -- this validates one of the unique value propositions of \clai, viz. offering instantaneous and on-premise support for new adopters of cloud platforms.
We also report on {\bf adoption tolerance} by measuring and reporting interest in using \clai.
This was done via a variety of questions, most specifically represented in Figure~\ref{fig:results-aggregate}(f) and Figure~\ref{fig:results-aggregate}(g). The former 
shows that users do not want overt changes to the CLI that they know and love.
The latter -- Figure~\ref{fig:results-aggregate}(g) -- talks to users' patience with processing time and latency in general. Users are not willing to tolerate latencies of more than $3$ seconds; however, there is a sizeable contingent of respondents who are happy to trade-off some latency for AI-enabled assistance.

\section{Looking Forward}



Project \clai was open sourced a few months ago at AAAI 2020,
to widespread interest in the AI community and in popular media (for full details, please see the CLAI wiki: \textcolor{blue}{\url{github.com/ibm/clai/wiki}}). 
It already has 180+ stars and the first few open-source
contributors. One of the most prolific open-source contributors 
is a co-author of this whitepaper.

\vspace{-5pt}
\paragraph{The \nlcmd Competition at NeurIPS 2020.}

One of the immediate impacts of Project \clai\ is the organization of competitions around the key user interaction
patterns discussed in this paper. Primary among them is  
\nlcmd @ NeurIPS 2020 (\textcolor{blue}{\url{ibm.biz/nlc2cmd}})
which revolves around translating natural language 
descriptions of command line tasks to their correct Bash syntax. 
Other competitions, such as around the automation use case
of observing and learning from user activity on the command line -- 
centered around AI planning and reinforcement learning techniques -- 
are planned.


\clearpage
\section*{Code / Data Share Notice}

In support of democratizing access to the command line, we provide links to the \clai\ system, \clai codebase, and data used to report the internal evaluation. These materials are provided for the community to reproduce and build on top of our work without any hindrances or hassles. 


\paragraph{Home} Project \clai\ home: \textcolor{blue}{\url{clai-home.mybluemix.net/}}

\paragraph{Code}

The code is open-source and is available at: 
\textcolor{blue}{\url{github.com/ibm/clai}}.



\paragraph{Data}

Aggregate anonymized data from the user study is available 
at: \textcolor{blue}{\url{ibm.biz/bb-survey-results}}.
The link is interactive: you can use this service to explore in further detail 
the differentiated results across different participant subgroups. 

\clearpage
\input{supple-whitepaper}

\clearpage
\bibliographystyle{plain}
\bibliography{neurips-whitepaper}

\end{document}

%% file: supple-whitepaper.tex
\setcounter{section}{0}
\clearpage

\section*{Appendix}
\subsection*{Project CLAI: Instrumenting the Command Line
as a New Environment for AI Agents}

The following supplementary material provides screenshots of all the user interaction 
patterns and \clai skills described in Section~\ref{action} in the paper.
The material also includes more details on the profiling results, and responses
from participants in the user study presented
in Section~\ref{yeti}.

\section{\clai Skills in Action}
\label{sec:skills_in_action}

\begin{figure}[!htp]
\centering
\includegraphics[width=\columnwidth]{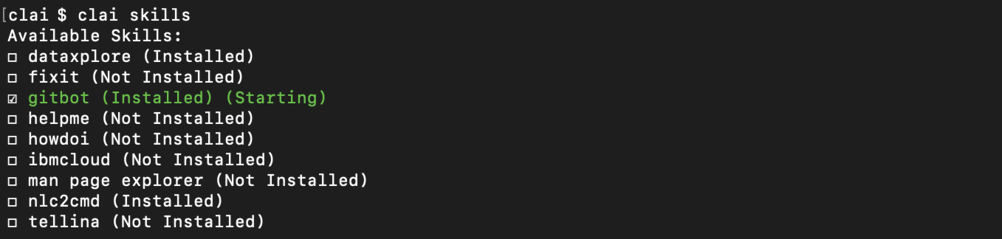}
\caption{
Screenshot of the \clai-enabled command line showing 
available active skills.}
\end{figure}

\begin{figure}[!htp]
\centering
\includegraphics[width=\columnwidth]{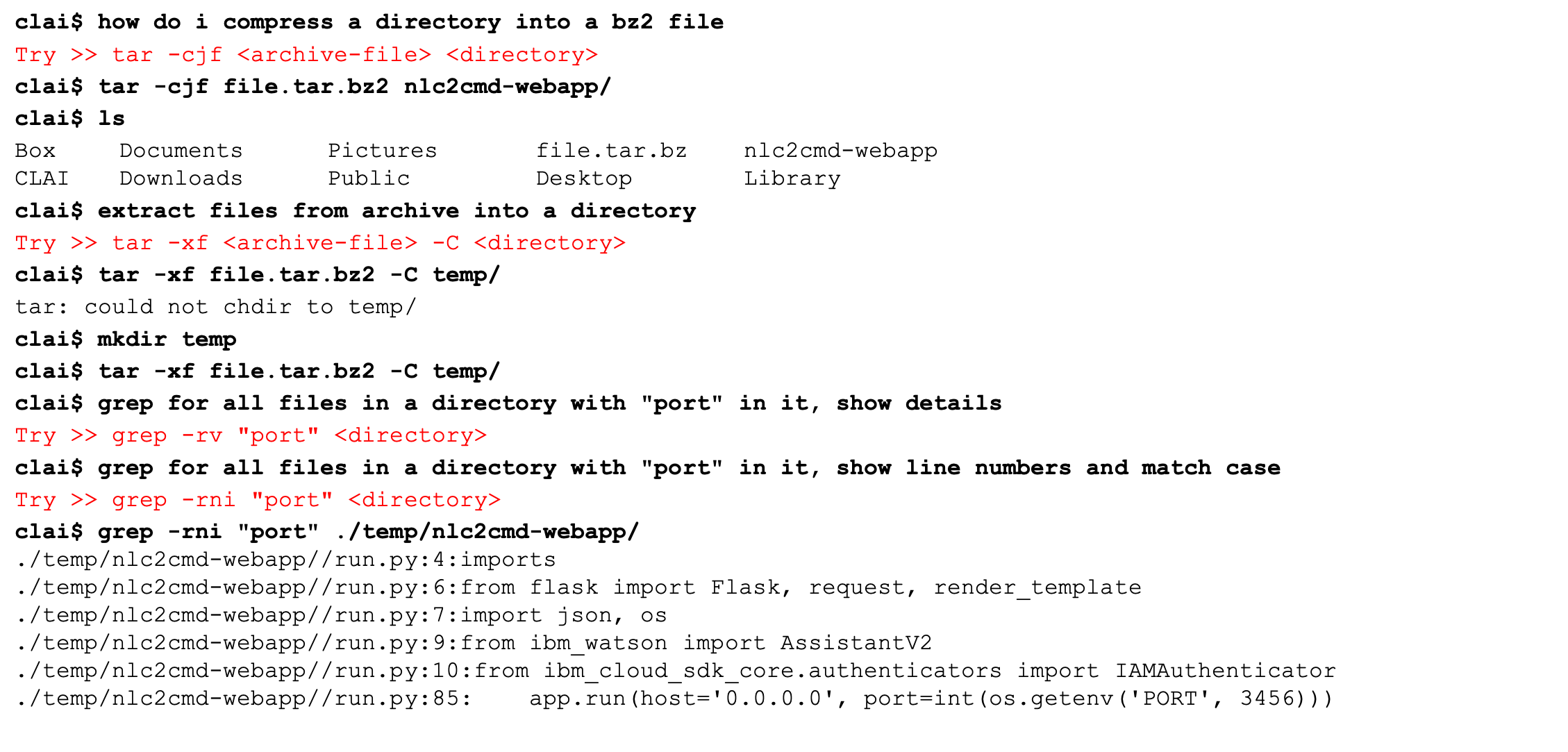}
\caption{
Screencast of a \clai-enabled command line illustrating how 
normal life on Bash is preserved unless the user invokes a skill
explicitly, or there is an event initiated by either
the user or the terminal itself where a \clai skill is confident 
about interjecting in.
Preserving the look and feel of Bash -- a tool near and dear to the
developer community -- is a crucial design choice we made 
based on the responses from the user study in Section \ref{study}; as well 
as lessons in the HCI community learned from previous deployments
\cite{cooper2004inmates}
of assistants for operating systems. 
}
\end{figure}

\clearpage

\subsection{nlc2cmd}

\begin{figure}[!htp]
\centering
\includegraphics[width=\columnwidth]{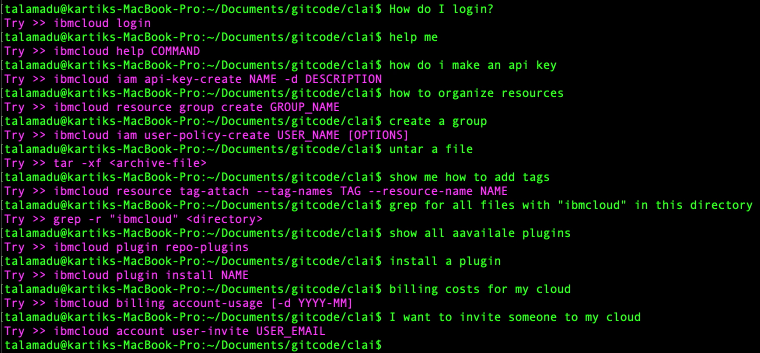}
\caption{
{\tt nlc2cmd} skill in action.
}
\label{fig:nlc2cmd-screen}
\end{figure}

\begin{center}
\begin{tabular}{rl}
{\bf Interaction Pattern} & On-premise support, natural language support \\
{\bf Supporting Technology} & Natural Language Processing
\end{tabular}
\end{center}

\subsection{tellina}

\begin{figure}[!htp]
\centering
\includegraphics[width=\columnwidth]{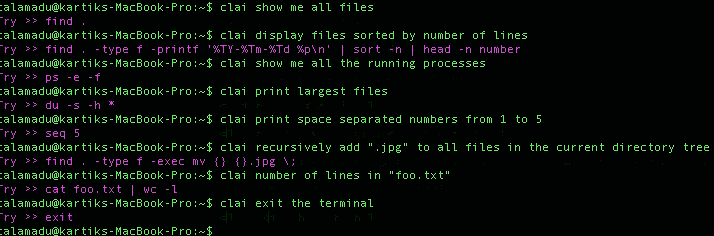}
\caption{
{\tt tellina} skill in action.
}
\label{invoke}
\label{fig:tellina-screen}
\end{figure}

\begin{center}
\begin{tabular}{rl}
{\bf Interaction Pattern} & On-premise support, natural language support \\
{\bf Supporting Technology} & Natural Language Processing
\end{tabular}
\end{center}

While {\tt nlc2cmd} (Fig.~\ref{fig:nlc2cmd-screen}) demonstrates more accurate (higher
inter-utility coverage) translation from English 
for specific commands (e.g. common bash utilities like 
{\tt tar} and {\tt grep} that trouble users the most \cite{lin2018nl2bash};
or platform specific commands like cloud services and mainframes),
{\tt tellina} (Fig.~\ref{fig:tellina-screen}) is the state of the art in general purpose
English to command translation, adopted from \cite{lin2018nl2bash}.

\clearpage

\subsection{fixit}

\begin{figure}[!htp]
\includegraphics[width=\columnwidth]{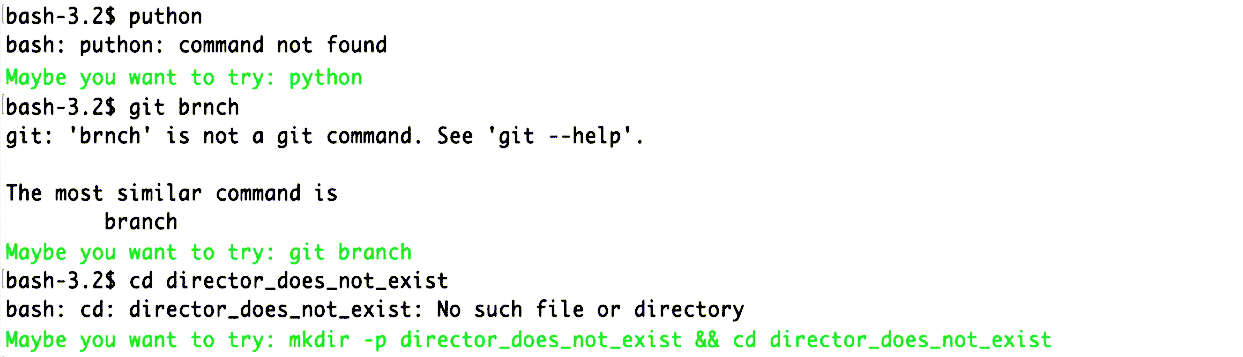}
\caption{
{\tt fixit} (demonstrating integration to Bash plugins like 
\textcolor{blue}{\url{github.com/nvbn/thefuck}})
}
\end{figure}

\begin{center}
\begin{tabular}{rl}
{\bf Interaction Pattern} & On-premise troubleshooting, proactive support\\
{\bf Supporting Technology} & Natural Language Processing, Automation
\end{tabular}
\end{center}

\subsection{cloudbot}

\begin{figure}[!htp]
\centering
\includegraphics[width=\columnwidth]{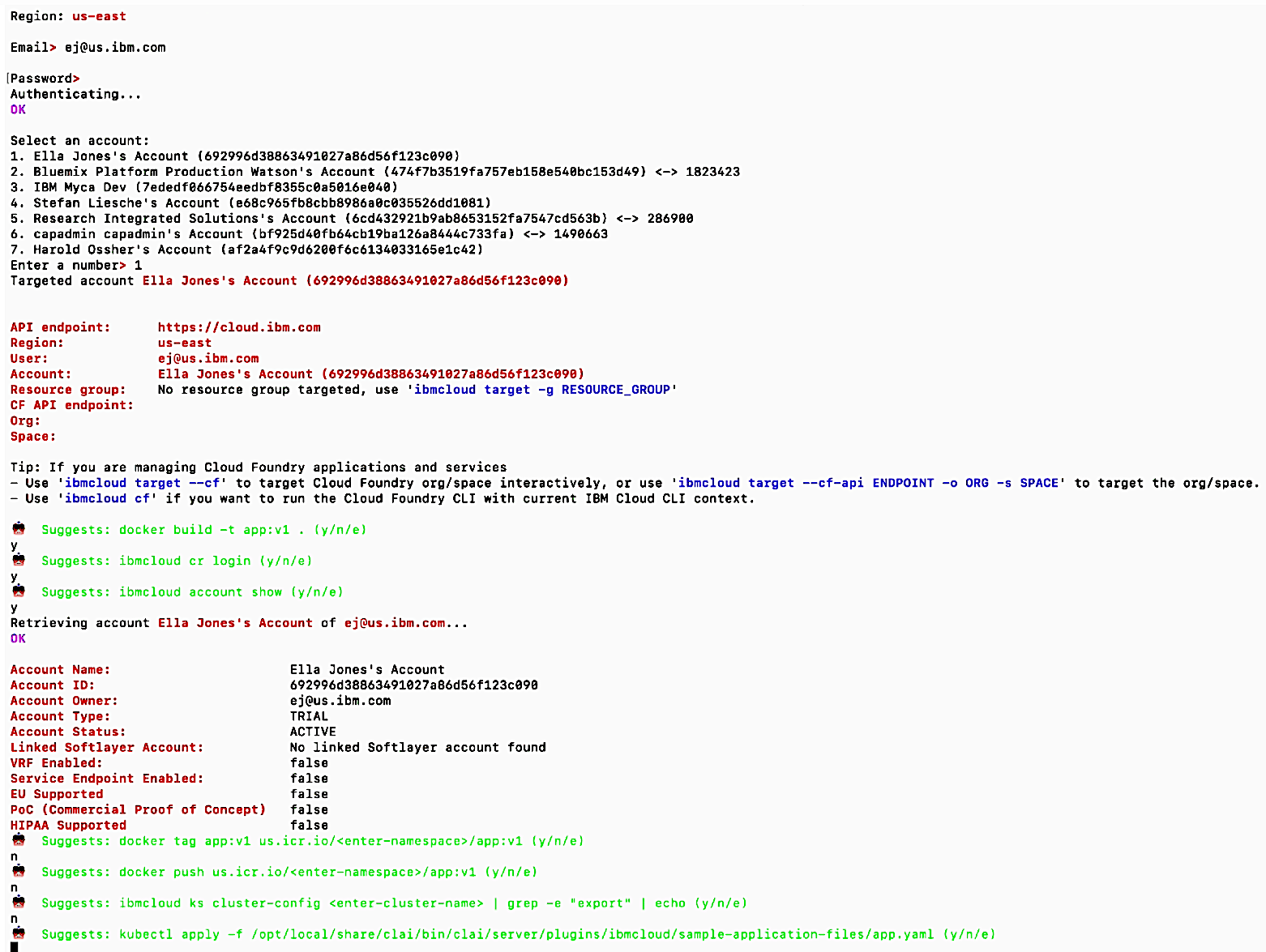}
\caption{
{\tt cloudbot} automation of the complex and lengthy deployment 
pipeline for a containerized application on to a Kubernetes platform.
Note that the user answers appear as ``no'' so as not to complete
the deployment in the course of taking screenshots. 
}

\end{figure}

\begin{center}
\begin{tabular}{rl}
{\bf Interaction Pattern} & Automation, On-premise support\\
{\bf Supporting Technology} & Planning 
\end{tabular}
\end{center}

\clearpage

\subsection{man page explorer}

\begin{figure}[!htp]
\includegraphics[width=\columnwidth]{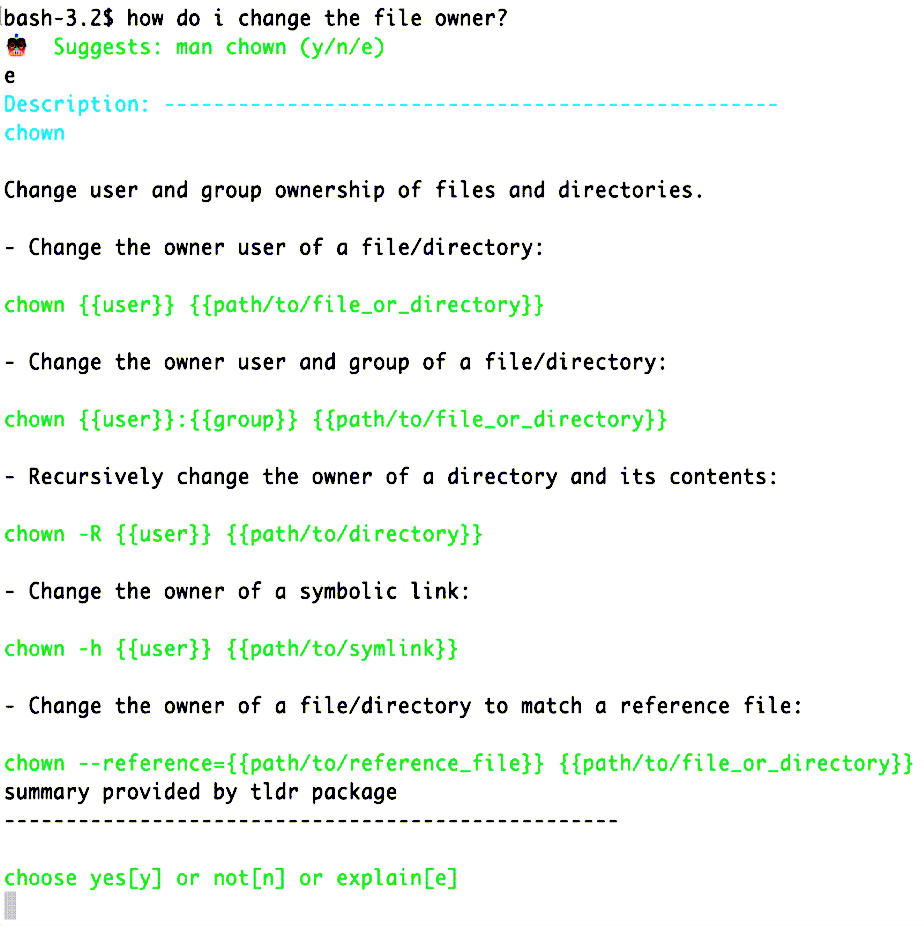}
\caption{
{\tt man page explorer} (demonstrating natural language support for
Q\&A functionality on the command line
along with integration to Bash plugins like 
\url{https://tldr.sh/}).
}
\end{figure}

\begin{center}
\begin{tabular}{rl}
{\bf Interaction Pattern} & On-premise support, natural language support,
augmentation\\
{\bf Supporting Technology} & Natural Language Processing, Q\&A and Summarization
\end{tabular}
\end{center}

This skill, along with {\tt howdoi} and {\tt helpme} described next,
are also examples where the user
uses the explain functionality described in the user controls 
in Section \ref{interaction-patterns} to gather more information 
on the command suggested by \clai.

\clearpage

\subsection{\clai howdoi}

\begin{figure}[!htp]
\includegraphics[width=\columnwidth]{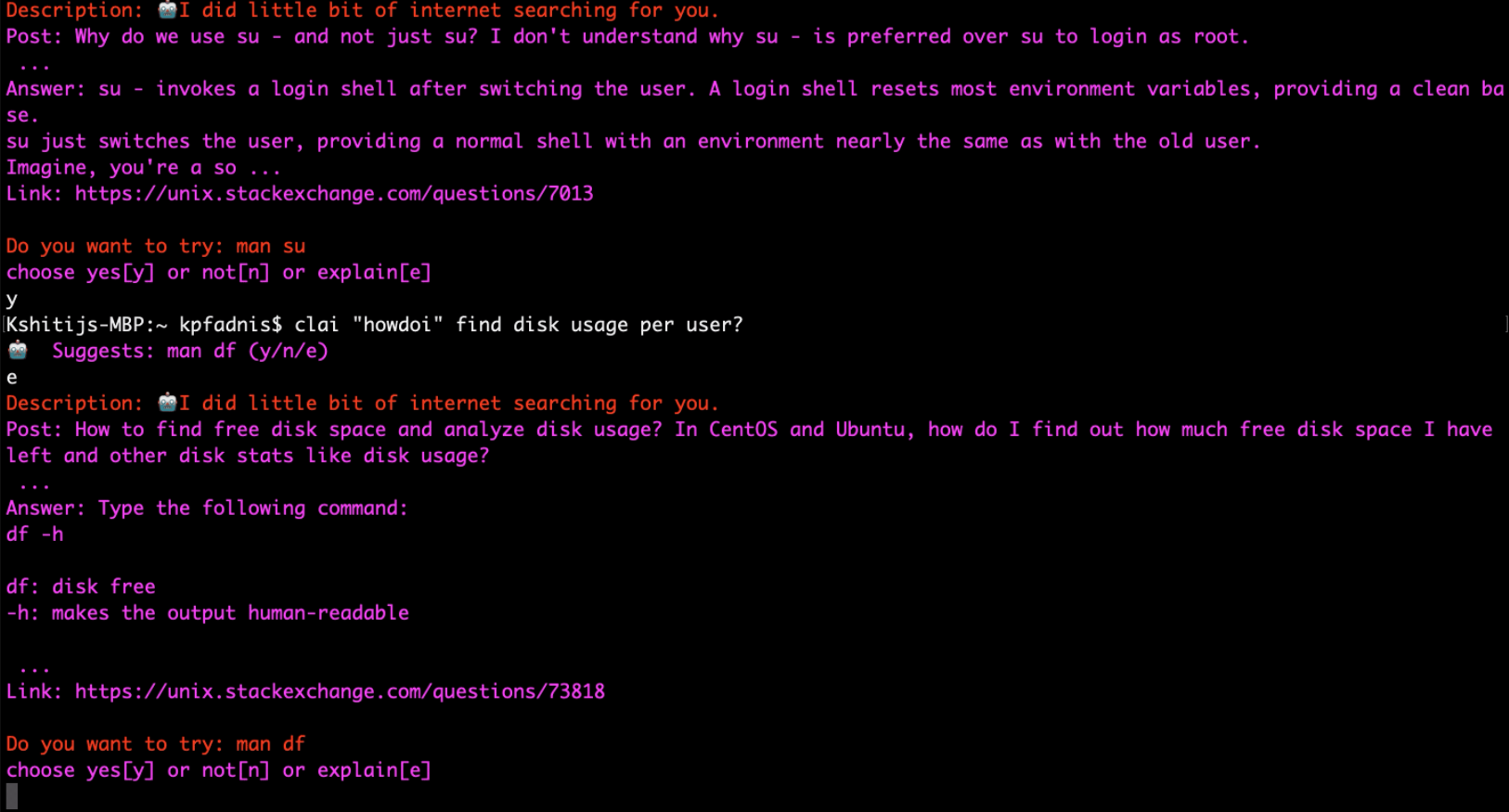}
\caption{
\clai\ {\tt howdoi} skill in action.
}
\end{figure}

\begin{center}
\begin{tabular}{rl}
{\bf Interaction Pattern} & On-premise support, natural language support,
augmentation\\
{\bf Supporting Technology} & Natural Language Processing, Information Retrieval
\end{tabular}
\end{center}

\subsection{\clai helpme}

\begin{figure}[!htp]
\includegraphics[width=\columnwidth]{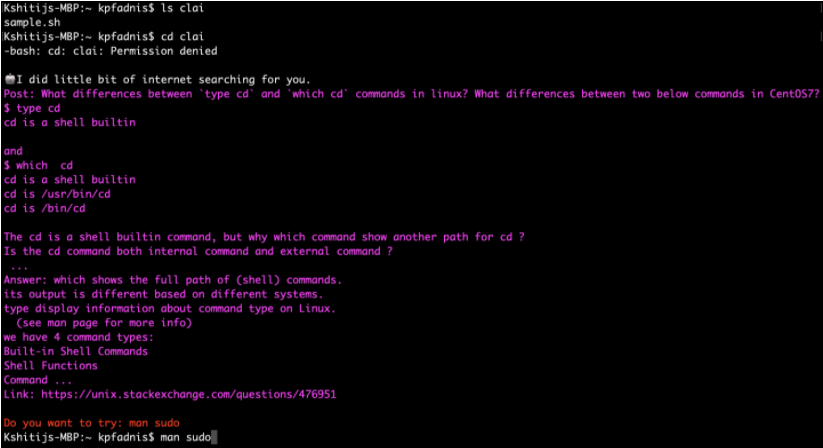}
\caption{
\clai\ {\tt helpme} skill in action.
}
\end{figure}

\begin{center}
\begin{tabular}{rl}
{\bf Interaction Pattern} & On-premise troubleshooting, proactive support,
augmentation\\
{\bf Supporting Technology} & Information Retrieval
\end{tabular}
\end{center}

\clearpage

\subsection{dataXplore}

\begin{figure}[!htp]
\centering
\includegraphics[width=\columnwidth]{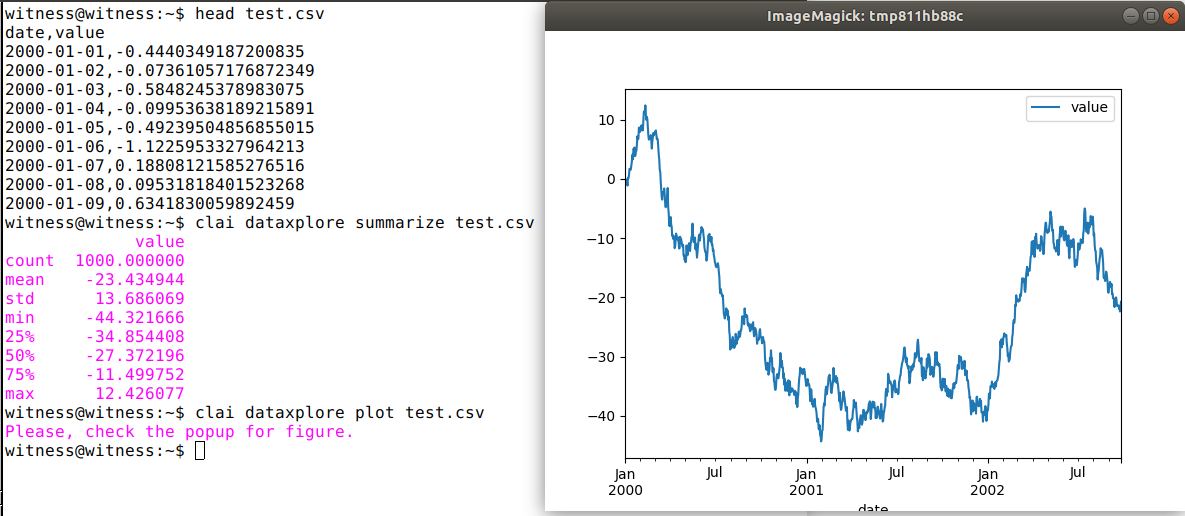}
\caption{
{\tt dataXplore} example of summarize after head function.
}

\end{figure}

\begin{figure}[!htp]
\centering
\includegraphics[width=\columnwidth]{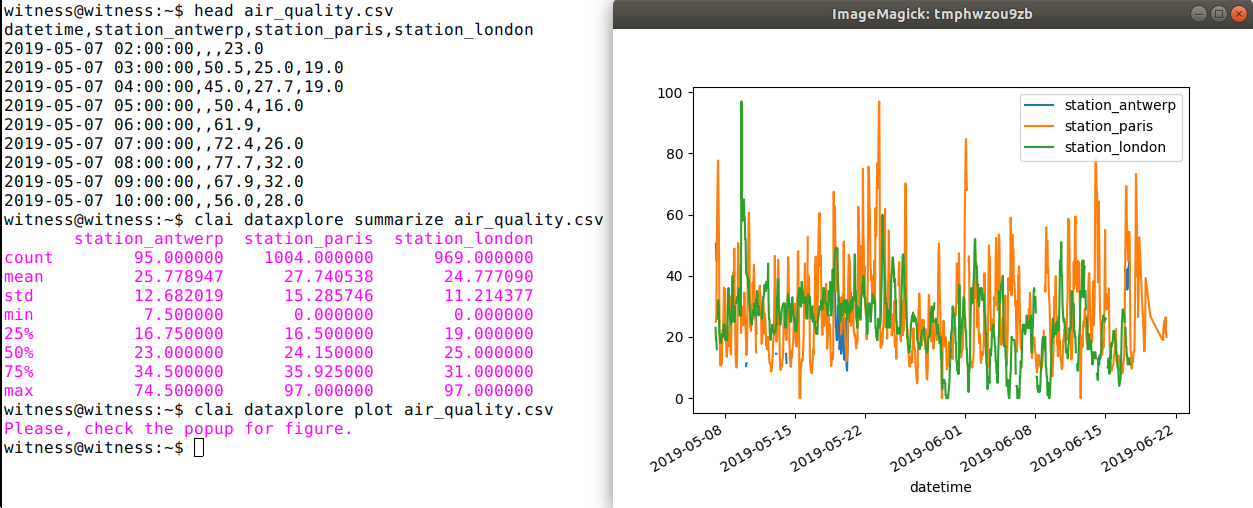}
\caption{
{\tt dataXplore} example of plot after head function.
}

\end{figure}

\begin{center}
\begin{tabular}{rl}
{\bf Interaction Pattern} & Augmentation\\
{\bf Supporting Technology} & Data Analysis
\end{tabular}
\end{center}

\clearpage

\subsection{gitbot}

\begin{figure}[!htp]
\centering
\includegraphics[width=\columnwidth]{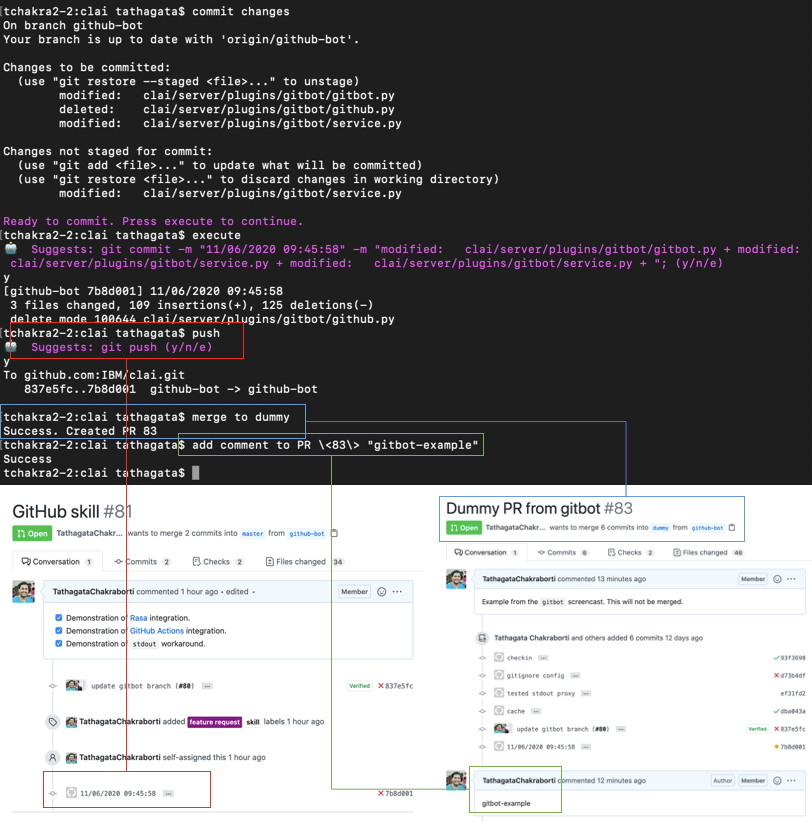}
\caption{
{\tt gitbot} example of local {\tt nlc2cmd} use case using
Rasa \cite{rasa} and remote GitHub operations using GitHub
Actions Developer API v3 \cite{github}. 
}

\end{figure}

\begin{center}
\begin{tabular}{rl}
{\bf Interaction Pattern} & Natural language support, automation\\
{\bf Supporting Technology} & Natural Language Processing
\end{tabular}
\end{center}

\clearpage

\section{\clai Orchestration in action}

All the above examples use the {\tt max-orchestration} pattern by 
default, as described in Section \ref{whale} -- notice how the user
needed to invoke \clai by force due to lower confidence of {\tt tellina}
in Figure \ref{invoke}, 
as per user controls introduced Section \ref{interaction-patterns}.
The following highlight the use of contextual bandits that can model
orchestration patterns specific to users and user types and can adapt
over time with continuous interaction, as described in Section 
\ref{whale}.

\subsection{Adaptive orchestration using contextual bandits}

\begin{figure}[!htp]
\centering
\includegraphics[width=\columnwidth]{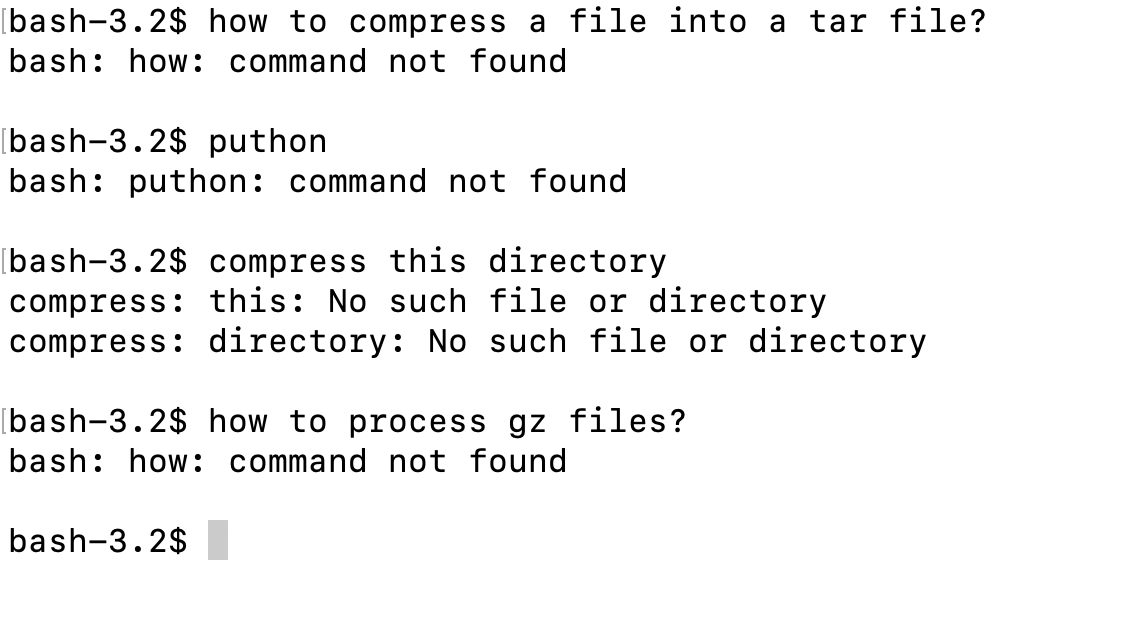}
\caption{``Ignore \clai'' orchestration behavior: the orchestrator here ignores any \clai skill response and treats each user command as a normal bash command. 
Notice how commands that usually elicit response from different skills (from previous snapshots) have been ignored and treated as native bash commands in the screenshot.
This profile can be used to ensure that normal bash commands are not hindered 
by \clai if the user has shown preference against it over time.
}
\label{fig:orchestration-ignore-clai}
\end{figure}

\begin{figure}[!htp]
\centering
\includegraphics[width=\columnwidth]{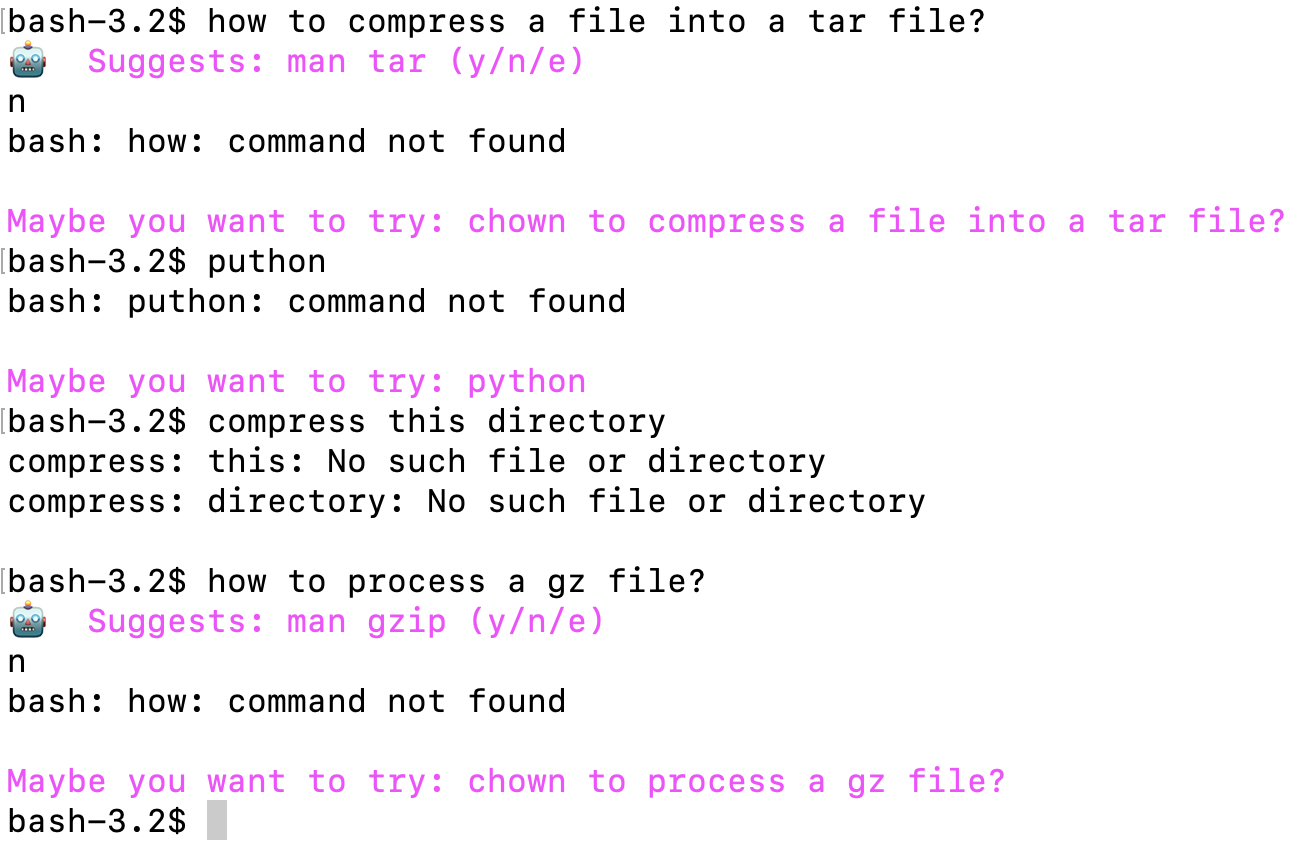}
\caption{``Ignore \texttt{nlc2cmd}'' skill orchestration behavior: 
the orchestrator here ignores response from the \texttt{nlc2cmd} skill and behaves as a \texttt{max-orchestrator} otherwise. Invocation 3 regarding compressing a directory would have elicited a response from the \texttt{nlc2cmd} skill, but is ignored in this warm-start profile.
This profile thus models individual preferences towards specific skills.}
\label{fig:orchestration-ignore-nlc2cmd}
\end{figure}

\begin{figure}[!htp]
\centering
\includegraphics[width=\columnwidth]{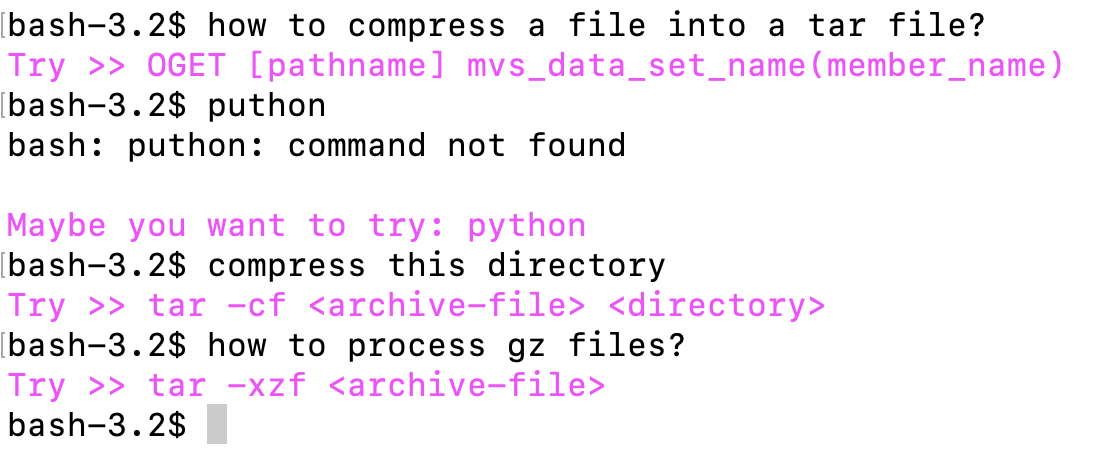}
\caption{``Maximum confidence'' orchestration behavior: the orchestrator here warm-starts with the behavior of a maximum confidence orchestrator, i.e., it selects the skill with the maximum confidence value. Note how the first invocaation is directed for the \texttt{man page explorer} skill but is responded to by the \texttt{nlc2cmd} skill because it has a higher confidence. This is a good default start behavior and the orchestrator can learn to adapt its behavior with the user feedback.}
\label{fig:orchestration-max-orchestrator}
\end{figure}

\begin{figure}[!htp]
\centering
\includegraphics[width=\columnwidth]{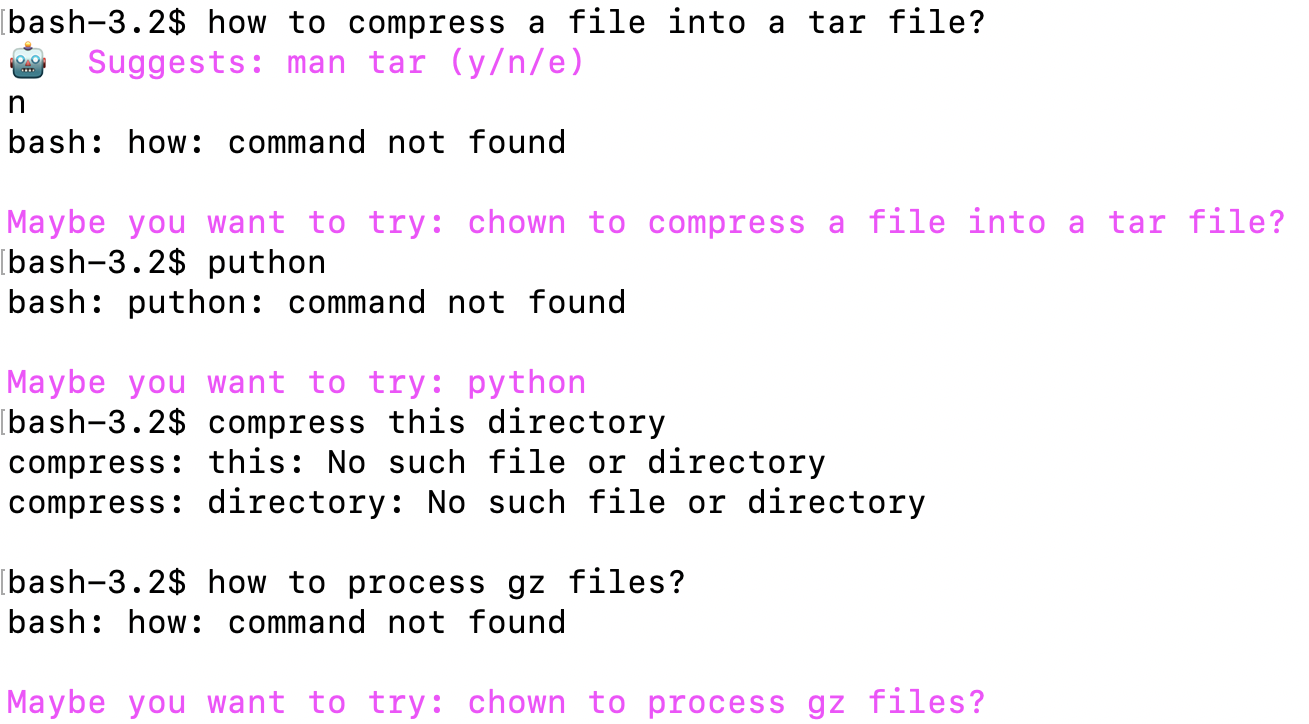}
\caption{``Prefer \texttt{man page explorer} skill over \texttt{nlc2cmd}'' orchestration behavior. The orchestrator here prefers the \texttt{man page explorer} skill over \texttt{nlc2cmd} and behaves as a maximum confidence orchestrator otherwise. Note that for the first invocation, a maximum confidence orchestrator would've chosen the \texttt{nlc2cmd} skill but since we prefer the \texttt{man page explorer} skill over the \texttt{nlc2cmd}, response from the \texttt{man page explorer} is chosen.
This is especially useful to model user types: e.g. a user on Mac may not 
find responses from Unix Stack Exchange useful and thus the bandit can learn
to prefer answers from man pages instead.}
\label{fig:orchestration-prefer-skill}
\end{figure}

The contextual bandit based orchestrator continues to learn and adapt its 
behavior depending on the feedback it receives from the user for each skill choice.
However, besides this online learning, these bandit orchestrators can also be warm-started
with a particular behavior profile to bypass the initial exploration phase of 
reinforcement learning algorithms. We demonstrate four different warm-start 
behavior patterns in figures \ref{fig:orchestration-ignore-clai}, \ref{fig:orchestration-ignore-nlc2cmd}, \ref{fig:orchestration-max-orchestrator}, and \ref{fig:orchestration-prefer-skill}.

\clearpage

\clearpage

\section{\clai System Footprint / Detailed Profiling Results}
\label{sec:system_footprint_profiling}


\begin{figure}[!htp]
\centering
\includegraphics[width=\textwidth]{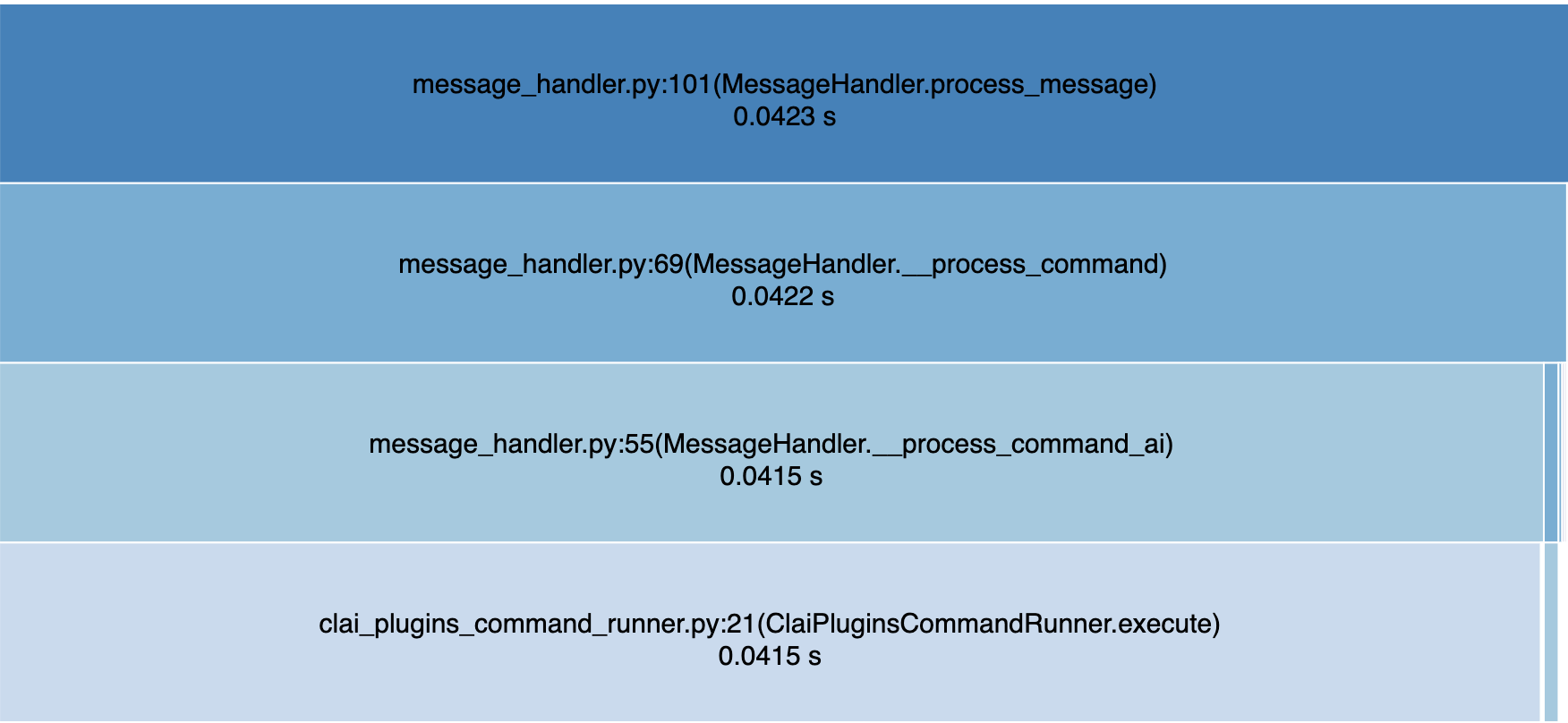}
\caption{Latency profile of the command \texttt{clai skills}. This invokes only the \clai core and thus provides a latency profile of \clai without the overhead of the individual skills.}
\label{fig:latency-clai-skills}
\end{figure}

\clearpage

\begin{figure}[!htp]
\centering
\includegraphics[width=\textwidth]{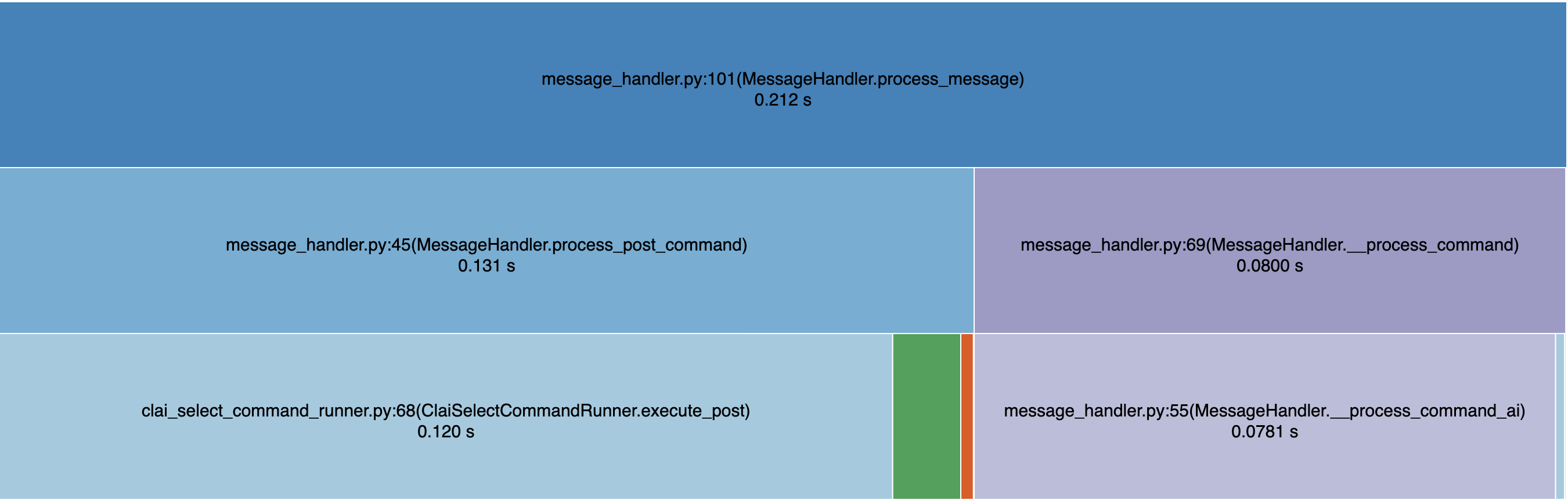}
\caption{Latency profile of activating a \clai skill. This activates the \clai core along with the activation procedure of the specified skill, thus increasing the latency but still keeping it well below the preferred sub-second mark.}
\label{fig:latency-clai-activate}
\end{figure}

\begin{figure}[!htp]
\centering
\includegraphics[width=\textwidth]{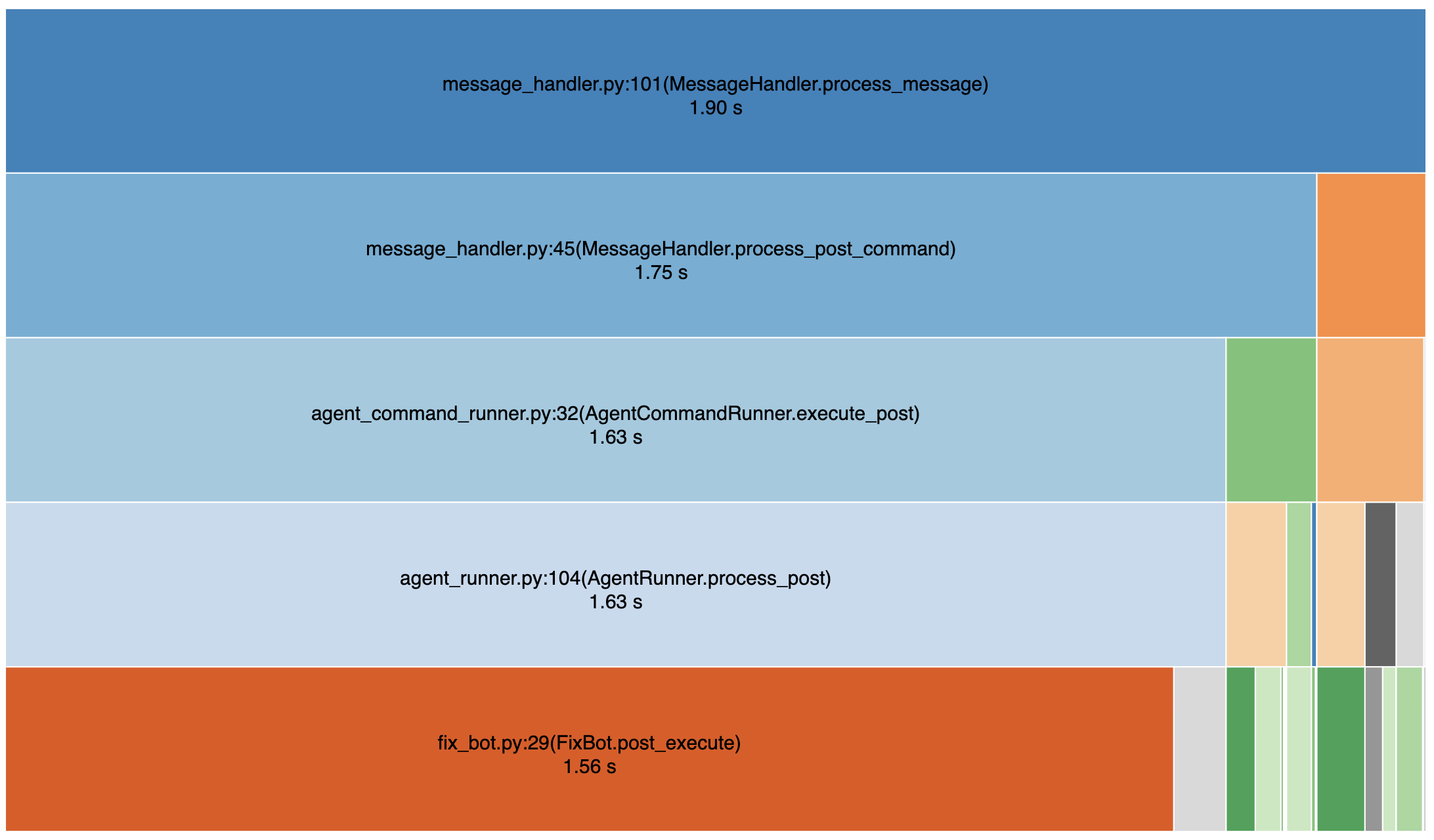}
\caption{Latency profile of executing a bash command with installed \texttt{clai skills}. This activates the \clai core along with \clai skills. While the latency increases significantly to 2 seconds, around 80\% of the latency is contributed by the skill itself.}
\label{fig:latency-clai-clearscreen}
\end{figure}

\section{\clai User Study}
\label{study}

\begin{figure}[!htp]
    \centering
    \includegraphics[width=1.0\linewidth]{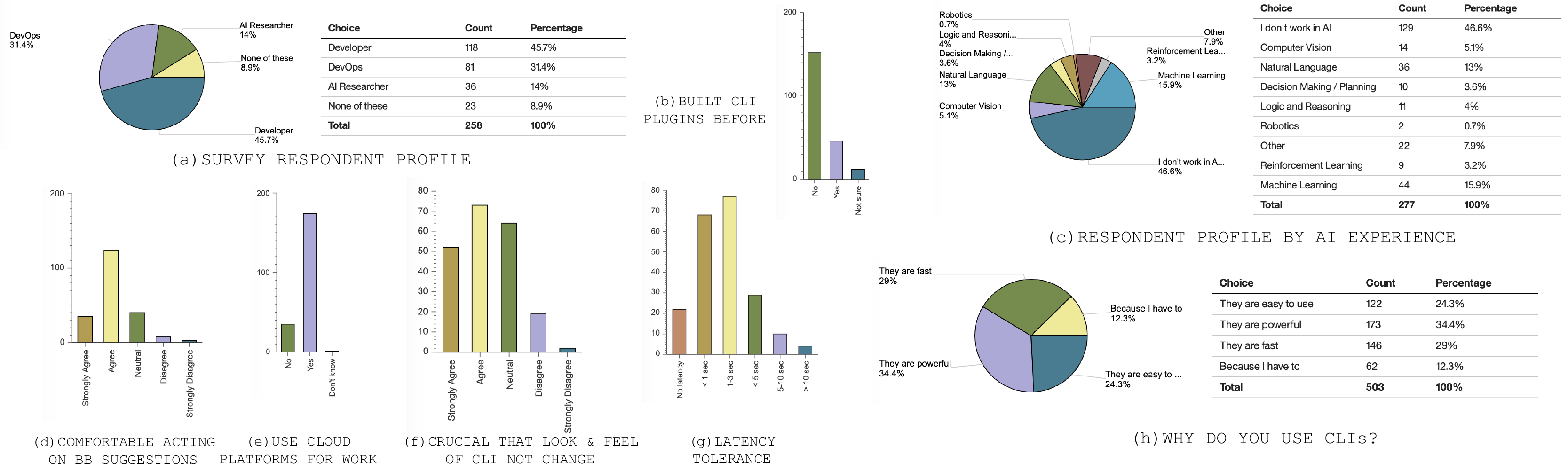}
    \caption{Combined \clai\ user study results, based on \numresponses\ responses.}
    \label{fig:results-combined-supplementary}
\end{figure}

In this section, we present an elaborated report of user feedback on \clai. These results are based on an internal survey of \numresponses\ respondents\footnote{An important note here that applies to our entire study is that although these results are based on \numresponses\ submitted responses, many of the questions allowed for the selection of multiple options -- hence some of the aggregate numbers are greater than the overall number of respondents.}. Figure~\ref{fig:results-combined-supplementary} provides an overall snapshot of the user study results, and is part of our main submission. In this section, we elucidate the details of the various sub-graphs within this figure, and analyze the aggregate responses. This section thus supplements the results presented in our ``Internal Evaluations'' section. We first detail the mechanism via which the survey was constructed and conducted; and then examine the results in detail. 

\subsection{Construction of User Survey}
\label{subsec:study_construction-survey}


We first detail the construction, administration, and deployment of the user survey. The survey was first storyboarded and assembled for coverage of all aspects of \clai\ that we wanted to measure and report on. Subsequent to this, the survey was deployed via the forms tool of a major multinational\footnote{Name omitted for double-blind review purposes.} software, services, and information technology company. The survey was sent out via email and other communication channels to a wide cross-section of employees, in the hopes of attracting responses from a diverse cohort of respondents who use or would have reason to use the command line interface.



The survey itself consisted of questions spread out over {\em three} main sections: {\em user demographics}, {\em interface patterns}, and {\em contributor questions}. The user demographics section collected information pertaining to a respondent's job role; the usage of cloud platforms and command line interfaces (CLIs) for their work; and specific information on the CLIs that they used. The interface patterns section surveyed respondents on the various ways that they could interface with \clai, and whether they thought that a specific pattern or feature would be useful to them. It additionally also surveyed users on extenuating factors like the latency introduced by an assistant like \clai, and how likely they were to use a CLI integrated with \clai. Finally, the contributor questions section surveyed whether the respondent would be interested in contributing to Project \clai; which branch of AI respondents had familiarity with (if any); and what new and upcoming features they would like to see in \clai. In the below, we collate and analyze the responses to some of these questions.

\subsection{Respondent Background \& Demographics}
\label{subsec:study_respondent-bg}

\begin{figure}[h]
    \centering
    \includegraphics[width=1.0\linewidth]{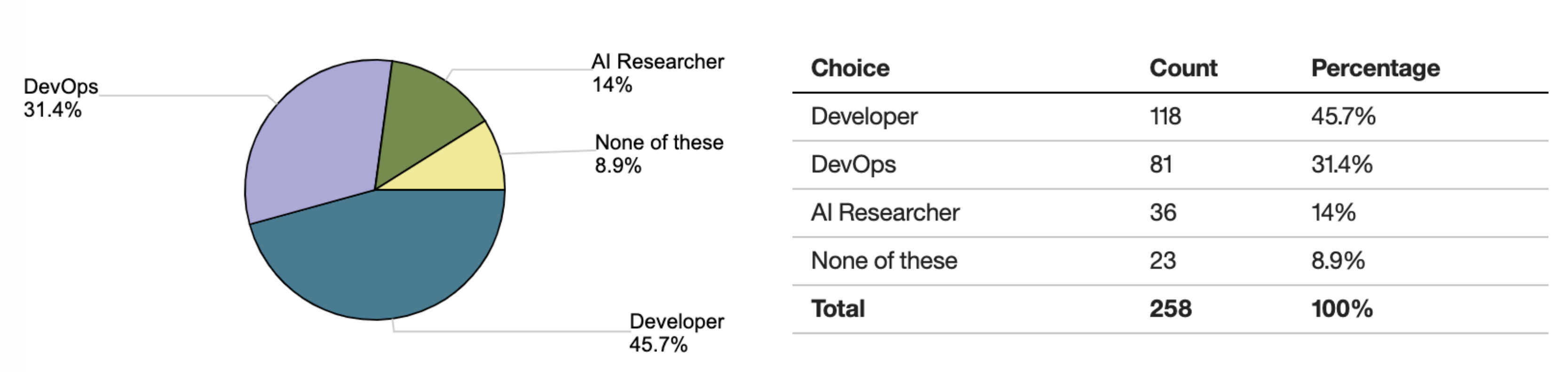}
    \caption{Demographics for user study respondents.}
    \label{fig:results-demographics-respondents}
\end{figure}

Figure~\ref{fig:results-demographics-respondents} profiles the survey respondents in aggregate. More than three-quarters of the responses came from respondents who identified as either developers or devops; while only around $14\%$ of the respondents identified as AI practitioners. 
This is an important insight, and shows the potential for \clai\ to positively impact communities that have hitherto not had too much interaction with state-of-the-art AI techniques and technologies.  This is a point that we elaborate on in our statement on the democratization of compute (included as part of the impact statement with the main paper). One of the main roles that we envision for \clai\ is as a vehicle for cutting-edge AI and ML techniques and tools to reach CLI users; and to break AI disciplines out of the cycle of validation and fine-tuning on toy domains and synthetic data (and datasets) alone. 



\subsection{AI Background} 
\label{subsec:study_ai-background}

\begin{figure}[htp]
    \centering
    \includegraphics[width=1.0\linewidth]{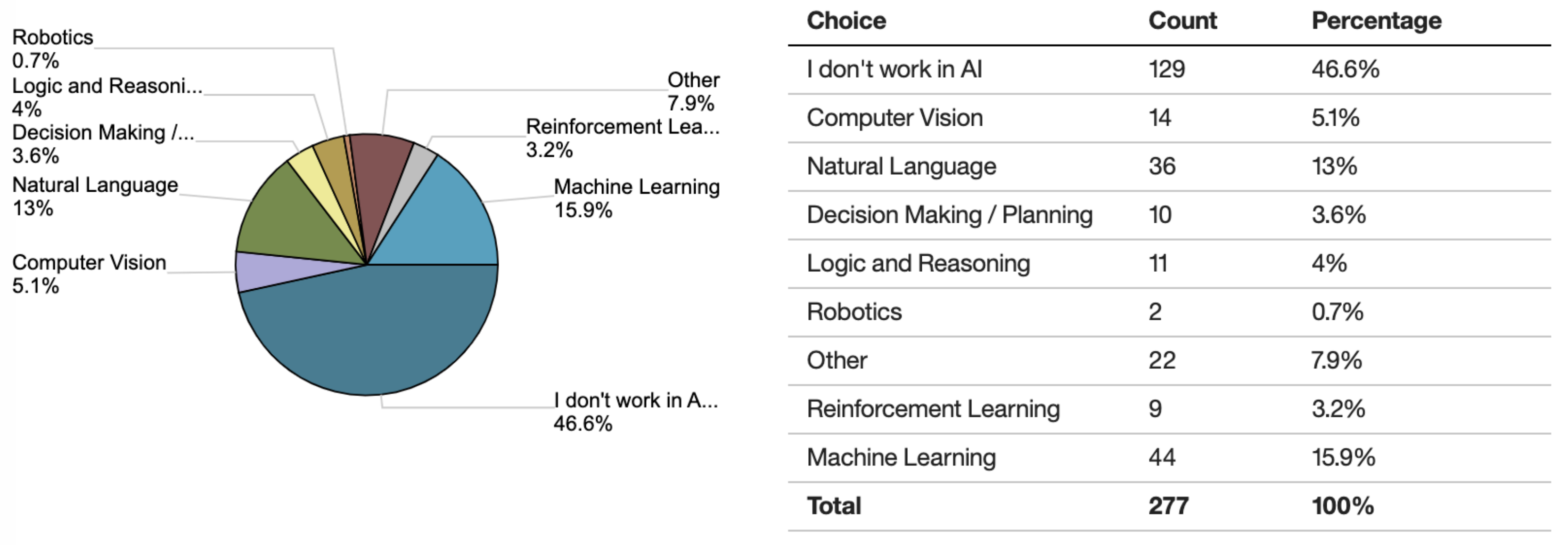}
    \caption{User study respondents' AI background information.}
    \label{fig:results-ai-background-respondents}
\end{figure}

Since \clai\ is intended first and foremost as a means to harness the state of the art in AI tools and techniques towards assistance on the command line, we naturally logged the AI background of the survey respondents. Figure~\ref{fig:results-ai-background-respondents} zooms in on the respondents' AI interests. Roughly over half of the responses indicated some experience with AI areas. Of these, Machine Learning was the top area; with Natural Language (Processing) a close second. These two results are expected, and follow the general trend in the AI world today, particularly when it comes to the non-research population. These also informed our choice of deploying the {\tt tellina} skill (c.f. Section~\ref{sec:skills_in_action}), which combines the latest technologies in deep learning and natural language processing. However, it is also interesting to note that there is still a reasonable population of respondents who are familiar with {\em classical} AI disciplines such as Decision Making, Logic \& Reasoning, and Reinforcement Learning. This wide distribution across AI topics is heartening, as it indicates the potential for new skills that can address unique CLI use-cases and become part of \clai\ in the future.

\subsection{Journey to Cloud}
\label{subsec:study_jtc}

\begin{figure}[htp]
    \centering
    \includegraphics[width=1.0\linewidth]{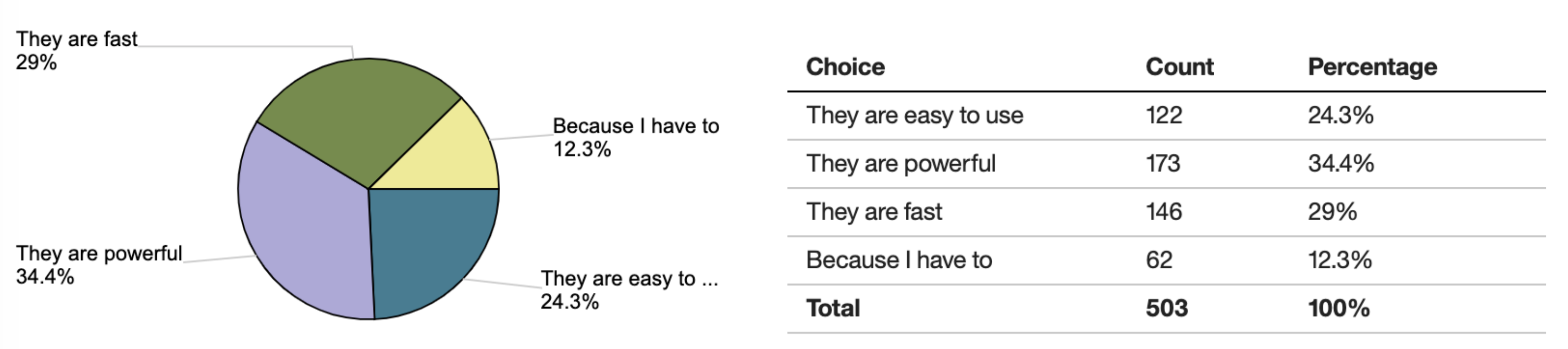}
    \caption{User study respondents' reasons for using cloud platforms.}
    \label{fig:results-cloud-respondents}
\end{figure}


In the introduction to this paper, we presented an argument for why the command line interface was making a resurgence; and why we are now at an inflection point that is akin to the initial emergence of large-scale networked terminals. Figure~\ref{fig:results-misc-subgraphs}(b) attests to this: an overwhelming majority of our respondents -- $4$ out of every $5$ -- report having to use cloud platforms for work. The reasons that respondents use cloud-based platforms are outlined in Figure~\ref{fig:results-cloud-respondents}: the major reasons mentioned by the respondents are all things that the CLI is known for. \clai\ is thus able to fulfil its unique value proposition of offering instantaneous and on-premise support for (new) adopters of cloud platforms. One of the ultimate goals of \clai\ -- reflected in our discussion on the various skills (c.f. Section~\ref{sec:skills_in_action}) -- is to ease users' {\em journey to the cloud} by making available various AI skills that are tuned toward specific use-cases; while always preserving the speed, power, and ease-of-use of the CLI.



\subsection{Adoption Tolerance}
\label{subsec:study_adoption-tolerance}

\begin{figure}[htp]
    \centering
    \includegraphics[width=1.0\linewidth]{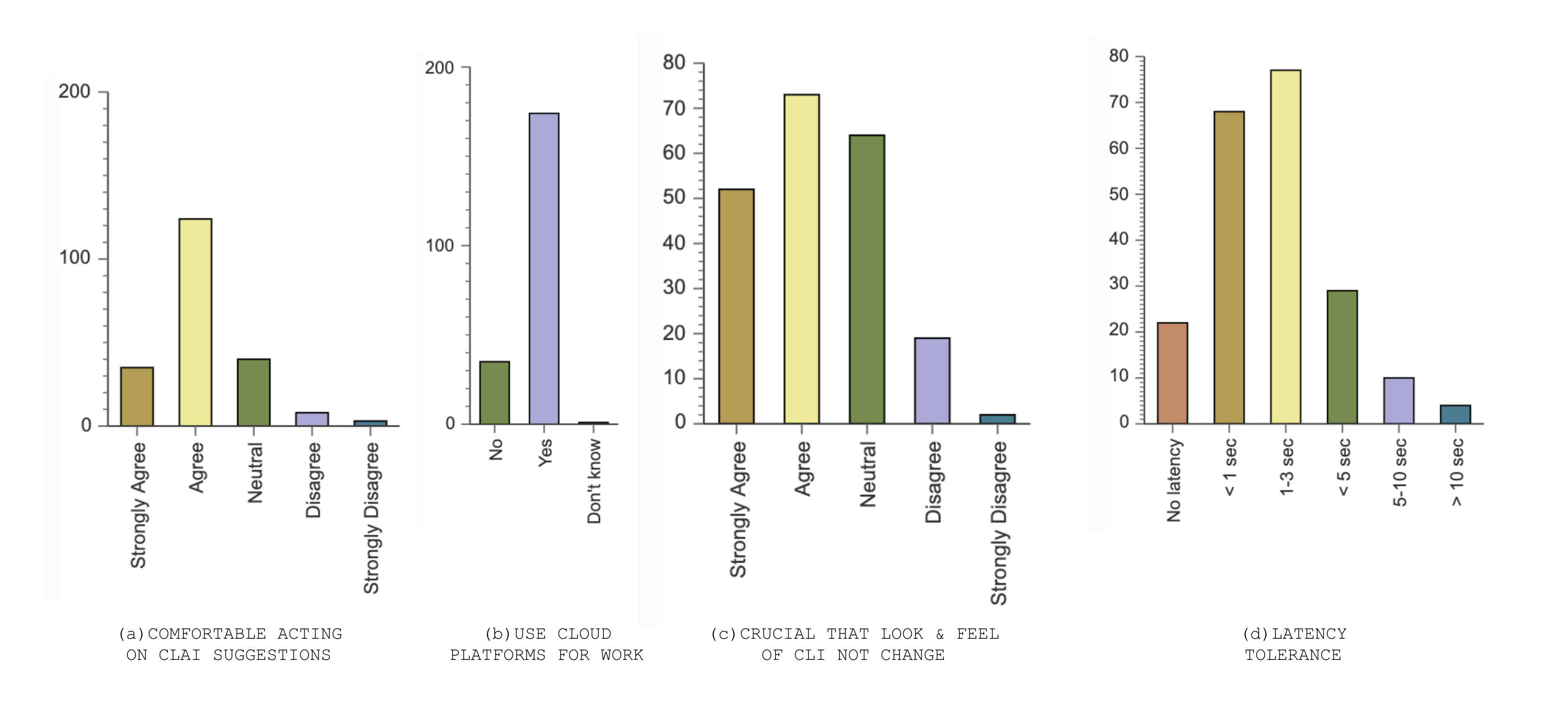}
    \caption{Results of users' responses on: (a) Willingness to act on \clai's suggestions; (b) Usage of cloud platforms for work; (c) Importance of look and feel of the CLI not changing; and (d) Latency tolerance.}
    \label{fig:results-misc-subgraphs}
\end{figure}

As with the introduction of any new technology, we measure and report the tolerance and appetite of the end-user for the tool being offered. This is particularly important in the context of smart assistants and assistive agents, which have had an unfortunate and infamous history of falling well short of user expectations~\cite{cooper2004inmates}. We measured the adoption tolerance of \clai\ via three questions, whose results are outlined in Figure~\ref{fig:results-misc-subgraphs}. The first -- Figure~\ref{fig:results-misc-subgraphs}(a) -- reports on users' willingness to act on suggestions from \clai. A significant majority of users feel comfortable acting on \clai's suggestions, which bodes well for adoption. Figure~\ref{fig:results-misc-subgraphs}(c) talks to the tendency of developers and other power-users to not want overt changes to the CLI that they know and love: indeed, a majority of respondents are disinclined to such changes. This validates our decision to provide the AI skills in \clai\ via the command line, as shown in the screen captures in Section~\ref{sec:skills_in_action}. Finally, we also measure users' tolerance to latency and processing times. Figure~\ref{fig:results-misc-subgraphs}(d) shows that users are not willing to tolerate latencies of more than 3 seconds; however, a sizeable contingent are happy to trade-off some increase in lantecy for the AI features provided by \clai. This validates our decisions and analyses while profiling \clai's system footprint (c.f. Section~\ref{sec:system_footprint_profiling}).

\subsection{Feature Usefulness}
\label{subsec:study_feature-usefulness}

\begin{figure}[!htp]
    \centering
    \includegraphics[width=1.0\linewidth]{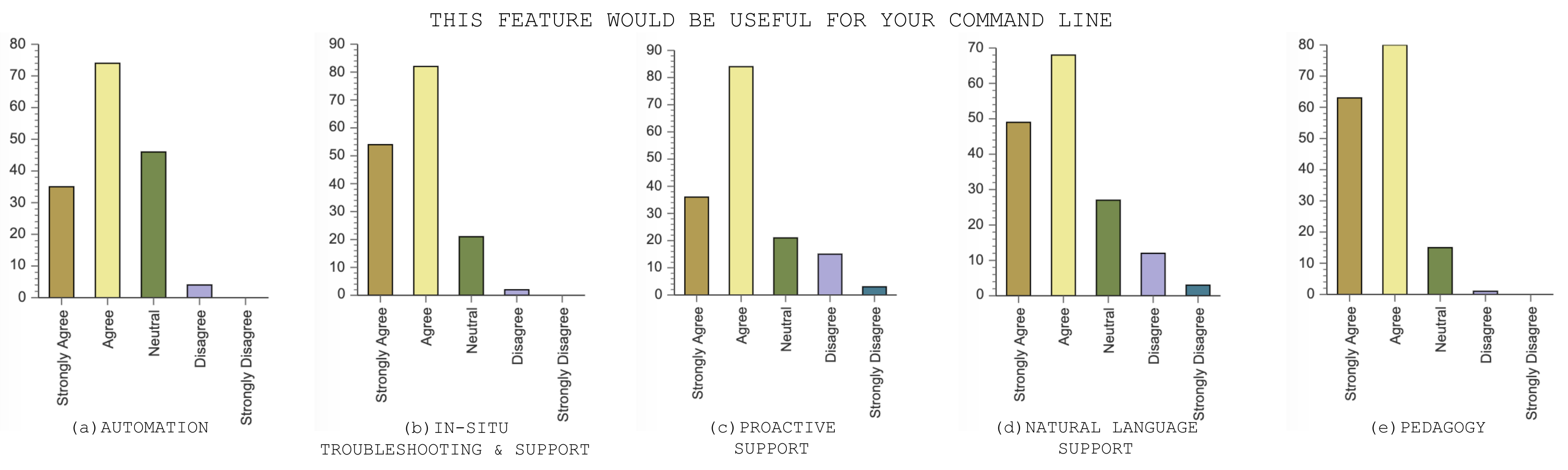}
    \caption{User study results for \clai\ features that users find useful on the CLI.}
    \label{fig:results-feature-would-be-useful}
\end{figure}

We also quizzed survey respondents on the various user interaction patterns and skills described in Section~\ref{interaction-patterns} of the paper -- these results are collated in Figure~\ref{fig:results-feature-would-be-useful}. There are some interesting trends. First, for the {\em Automation} pattern -- Figure~\ref{fig:results-feature-would-be-useful}(a) -- many users expressed a {\tt neutral} response. This may be attributed to the fact that automation of entire processes is the most complex assistance possible; and thus users are likely to be most wary of this. Users also expressed very little negative sentiment towards the {\em in-situ troubleshooting \& support} (Figure~\ref{fig:results-feature-would-be-useful}(b)) and {\em pedagogy} (Figure~\ref{fig:results-feature-would-be-useful}(e)) patterns. The former can be explained by the fact that this is currently the killer app that most users of the command line are waiting for: users do not want to interrupt their task to go look for solutions. The latter can be construed as an aspirational goal, where users are looking forward to a true command line assistant that can make them better and more efficient power-users with sustained usage. Finally, both the {\em proactive support} (Figure~\ref{fig:results-feature-would-be-useful}(c)) and {\em natural language support} (Figure~\ref{fig:results-feature-would-be-useful}(d)) use-cases had high positive sentiment, coupled with a bit of user anxiety about how these patterns would be implemented as skills. We take these results to indicate that user adoption of these patterns comes down to their specific implementation as skills. 


\subsection{Using \& Extending \clai}
\label{subsec:study_using-extending-clai}

\begin{figure}[htp]
    \centering
    \includegraphics[width=0.85\linewidth]{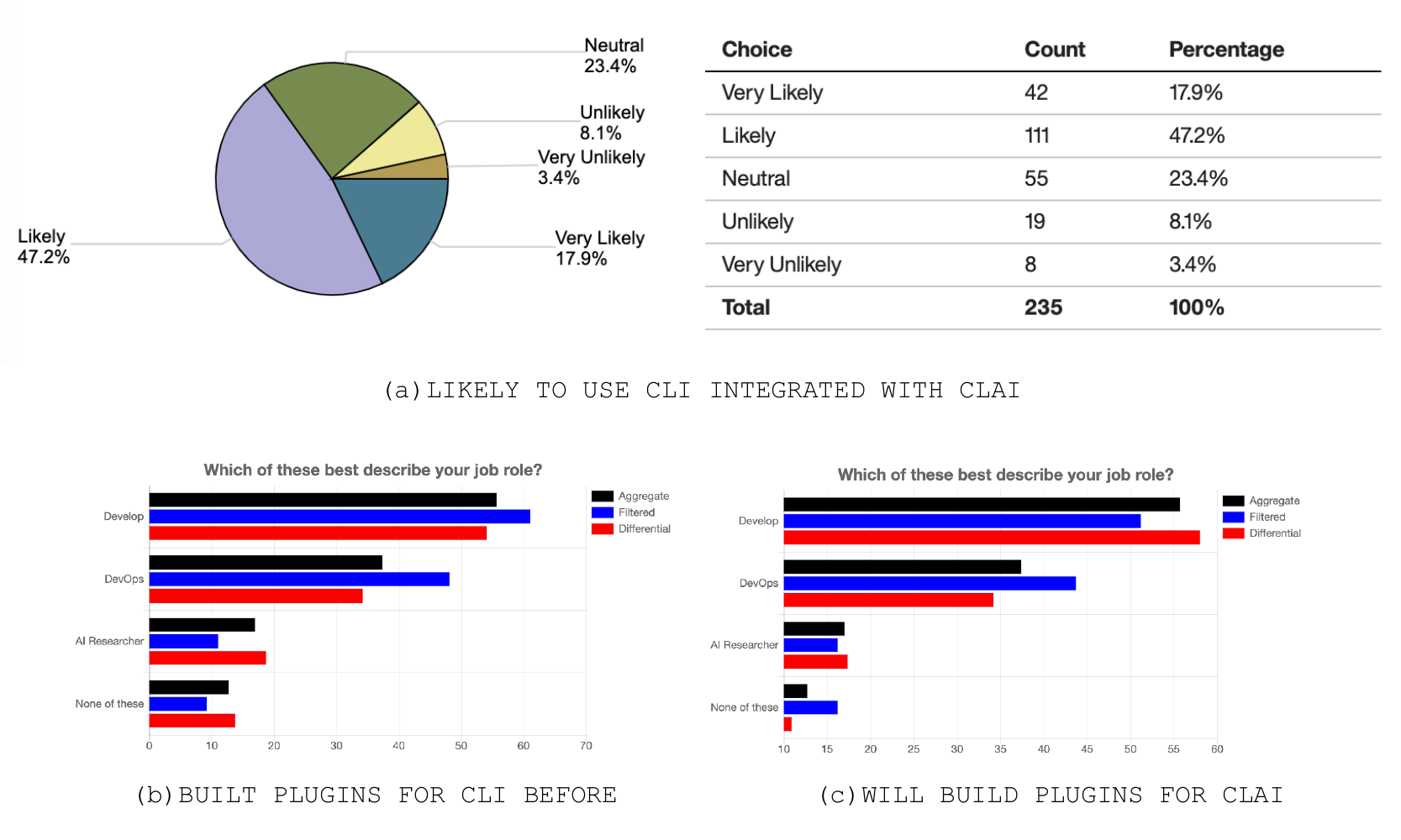}
    \caption{User study results on usability of \clai.}
    \label{fig:results-usability-clai-platform}
\end{figure}


Finally, we surveyed users on whether they would use a version of the CLI integrated with \clai\ -- these results are aggregated in Figure~\ref{fig:results-usability-clai-platform}(a). The vast majority of users report that they are likely to use \clai. 
Furthermore, 
we present Figure~\ref{fig:results-usability-clai-platform}(b) and Figure~\ref{fig:results-usability-clai-platform}(c), which respectively show the breakdown by role of respondents who have built plugins for the CLI {\em previously}, and those who would build {\em new} AI-based plugins for \clai. It is particularly informative to note the difference in numbers between the {\em developer} and {\em researcher} job roles.
